\documentclass[letterpaper,11pt]{article}

\usepackage{jheppub}

\newcommand{\eq}[1]{eq.~\eqref{eq:#1}}
\newcommand{\eqs}[2]{eqs.~\eqref{eq:#1} and \eqref{eq:#2}}
\renewcommand{\sec}[1]{sec.~\ref{sec:#1}}

\newcommand{\app}[1]{app.~\ref{app:#1}} 
\newcommand{\fig}[1]{fig.~\ref{fig:#1}}

\newcommand{\ord}[1]{{\mathcal O}(#1)}
\newcommand{\ORd}[1]{{\mathcal O}\Bigl(#1\Bigr)}

\newcommand{\nn}{\nonumber}

\newcommand{\vecb}[1]{\mbox{\boldmath $#1$}}

\def\({\left(}
\def\[{\left[}
\def\){\right)}
\def\]{\right]}

\newcommand{\df}{\mathrm{d}}
\newcommand{\sdt}{\!\cdot\!}

\newcommand{\al}{\alpha}

\newcommand{\ga}{\gamma}

\newcommand{\de}{\delta}

\newcommand{\la}{\lambda}
\newcommand{\si}{\sigma}

\newcommand{\cG}{{\mathcal G}}
\newcommand{\cJ}{{\mathcal J}}

\newcommand{\bn}{{\bar{n}}}
\newcommand{\nbar}{\bar{n}}

\newcommand{\lqcd}{\Lambda_\mathrm{QCD}}

\newcommand{\zero}{{(0)}}
\newcommand{\one}{{(1)}}

\allowdisplaybreaks[2]

\title{Jet axes and universal transverse-momentum-dependent fragmentation}

\author[a]{Duff Neill,}
\author[b]{Ignazio Scimemi,}
\author[c,d]{Wouter J.~Waalewijn}

\affiliation[a]{Theoretical Division, MS B283, Los Alamos National Laboratory, Los Alamos, NM 87545, USA}
\affiliation[b]{Departamento de F\'isica Te\'orica II, Universidad Complutense de Madrid, Ciudad Universitaria, 28040 Madrid, Spain}
\affiliation[c]{Institute for Theoretical Physics Amsterdam and Delta Institute for Theoretical Physics, University of Amsterdam, Science Park 904, 1098 XH Amsterdam, The Netherlands}
\affiliation[d]{Nikhef, Theory Group, Science Park 105, 1098 XG, Amsterdam, The Netherlands}

\emailAdd{duff.neill@gmail.com}
\emailAdd{ignazios@fis.ucm.es}
\emailAdd{w.j.waalewijn@uva.nl}

\abstract{
We study the transverse momentum spectrum of hadrons in jets. By measuring the transverse momentum with respect to a judiciously chosen axis, we find that this observable is insensitive to (the recoil of) soft radiation. Furthermore, for small transverse momenta we show that the effects of the jet boundary factorize, leading to a new transverse-momentum-dependent (TMD) fragmentation function. In contrast to the usual TMD fragmentation functions, it does not involve rapidity divergences and  is universal in the sense that it is  independent of the type of process and number of jets. These results directly apply to sub-jets instead of hadrons. We discuss potential applications, which include studying nuclear modification effects in heavy-ion collisions and identifying boosted heavy resonances.
}
\preprint{NIKHEF 2016-051}

\begin{document}
\maketitle

%%%%%%%%%%%%%%%%%%%%%%%%%%%%%%%%%%%%%%%%%%%%%%%%%%%%%%%%%%%%%%%%%%%%%%%%%%%%%%%%
\section{Introduction}
\label{sec:intro}
%%%%%%%%%%%%%%%%%%%%%%%%%%%%%%%%%%%%%%%%%%%%%%%%%%%%%%%%%%%%%%%%%%%%%%%%%%%%%%%%

In the analysis of events from hadron colliders it is common to use jets to organize the final states of hard interactions, making it natural to ask how the QCD confinement of hadrons is realized in this context.  
The picture that arises from QCD factorization is that we have the hard scattering, whose calculation is given in terms of partonic degrees of freedom, initiating the jet. At the short-distance scale of the hard-scattering, we have a quark or gluon of a much lower ``off-shellness'' exiting the hard interaction in a more or less definite direction. The subsequent branching does not change this direction much, but does gives rise to a host of additional partons loosely grouped into a jet. These are the perturbative remains of the slightly off-shell parton. Lastly, these additional partons undergo a ``hadronization'' process at length scales of $1/\Lambda_{\rm QCD}$, confining themselves into the observed hadrons. Ultimately, to understand the dynamics of confinement within jets, we would like to have a means of comparing the partonically generated momentum distribution inside the jet to the observed hadronic momentum distribution. In addition to momentum, one would also like to understand how quantum numbers, like spin, flavor, or charge, are transported from the hard scattering into the hadronic final state. 

The fragmentation function $d_{i \to h}(z_h,\mu)$ describes the distribution of the longitudinal-momentum fraction $z_h$ of hadrons of a species $h=\pi^+, \pi^-, \dots$ produced by a parton $i=g, u, \bar u, d, \dots$~\cite{Georgi:1977mg,Ellis:1978ty,Collins:1981uw}. This allows one to express their production cross section as (see e.g.~ref.~\cite{Collins:1989gx})
%%%
\begin{align}
  \frac{\df \sigma_h}{\df z_h} = \sum_i \int\! \frac{\df z}{z}\, \hat \sigma_i(z,Q,\mu) \, d_{i \to h}\Big(\frac{z_h}{z},\mu\Big) \Big[1+ \ORd{\frac{\lqcd}{Q}}\Big]
,\end{align}
%%%
where $Q$ is the scale of the hard scattering.
A crucial feature of fragmentation is that it is universal, i.e.~insensitive to the underlying hard scattering or the soft background radiation. In field-theoretic terms this means that the same QCD matrix element for $d_{i \to h}$ captures the fragmentation dynamics, and can be factorized from the hard scattering.
Thus fragmentation measurements at hadron-hadron, hadron-electron, and electron-positron colliders can all be compared. 
However, when  combining hadron analysis with modern jet algorithms
one begins to worry that the definition of the jet itself could potentially spoil this universality, since any given jet definition will have more or less sensitivity to the underlying event or hard scattering process. As we will see in this paper this can take a rather subtle form.

Fragmentation of hadrons \emph{inside} jets has also been studied extensively, but without accounting for the transverse momentum dependence of the hadrons. When the jet is sufficiently narrow, its dynamics can be factorized from the hard scattering process. For fragmentation in exclusive processes (i.e.~with a specific number of jets) this was studied using event shapes (hemisphere jets) in refs.~\cite{Procura:2009vm, Liu:2010ng, Jain:2011xz, Jain:2012uq, Bauer:2013bza, Ritzmann:2014mka} and with a jet algorithm in refs.~\cite{Procura:2011aq,Chien:2015ctp,Baumgart:2014upa,Bain:2016clc}. Inclusive jet production with a jet algorithm was investigated in refs.~\cite{Arleo:2013tya,Kaufmann:2015hma,Dai:2016hzf,Kang:2016ehg}. The applications that were considered range from comparisons to LHC measurements of charged hadron spectra~\cite{Chien:2015ctp} to unravelling quarkonium production channels~\cite{Baumgart:2014upa}. Multi-hadron fragmentation in jets has also been considered~\cite{Krohn:2012fg,Waalewijn:2012sv,Chang:2013rca,Chang:2013iba}, to e.g.~describe jet charge~\cite{Krohn:2012fg}.

The observables that we want to construct here are transverse momentum distributions (TMDs). In general, one would like to know the full three-dimensional distribution of momenta inside the jet, not merely the energy fraction. However, one must be careful, since asking questions about the other components of the hadron's momentum can easily expose one to sensitivity to associated soft processes. While studying these soft processes is an interesting and worthwhile endeavor in and of itself, it can severely complicate any potential claim to universality of these distributions. In the standard terminology the TMDs measure the correlation of transverse momentum of two partons in processes like Drell-Yan, semi-inclusive deep-inelastic scattering (SIDIS), or the production of two hadrons in $e^+e^-$ collisions. In this work we consider instead the measurement of the transverse momentum of a hadron with respect to a jet axis. In the standard TMD correlations one cannot avoid the appearance of rapidity divergences and the consequent regularization and renormalization \cite{Ji:2004wu,Ji:2004xq,Becher:2010tm,Collins:2011zzd,Chiu:2012ir,GarciaEchevarria:2011rb}, which signals a sensitivity to soft physics.

\begin{figure}
\centering
\includegraphics[width=0.6\textwidth]{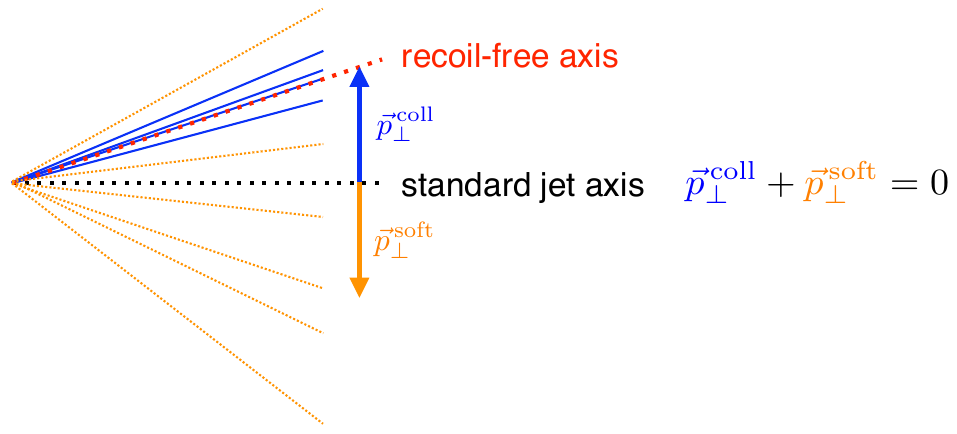}%
\caption{The standard jet axis is sensitive to soft radiation (orange) through recoil effects, whereas a recoil-free axis follows the direction of collinear radiation (blue).}
\label{fig:recoil}
\end{figure}

Here we show that one can define a transverse momentum observable which is insensitive to such problems, by a judicious choice of jet axis. The final TMDs will necessarily be different from the standard ones and  thus we coin the name  jet TMDs (JTMDs) for this class of observables. The key insight is to adopt an axis definition that is recoil-insensitive~\cite{Bertolini:2013iqa,Larkoski:2014uqa,Salam:WTAUnpublished}. Put loosely, if one uses an axis whose direction is conserved under splittings (e.g.~the total momentum of the jet or the thrust axis), this introduces a soft-sensitivity in the axis definition, since a soft emission can displace a collinear one, see \fig{recoil}. Alternatively, one can adopt an axis that itself recoils coherently with the soft radiation, and thus follows the direction of the collinear radiation.

In describing the transverse momentum of a hadron in a jet, there are a number of choices:
\begin{itemize}
\item Exclusive production with a jet algorithm with a recoil-sensitive axis: the factorization theorem has a simple multiplicative structure (see e.g.~ref.~\cite{Ellis:2010rwa}) but the soft radiation suffers from non-global logarithms (NGLs)~\cite{Dasgupta:2001sh}, which arise because of the very different restrictions on the radiation inside an outside the jet. 
\item Exclusive production with a global event shape: NGLs are absent for an observable like $N$-jettiness~\cite{Stewart:2010tn}, but potential exchanges between the initial states can spoil the factorization for hadronic collisions~\cite{Forshaw:2012bi,Rogers:2010dm,Catani:2011st,Gaunt:2014ska,Zeng:2015iba,Rothstein:2016bsq}.
\item 
Inclusive production with a jet algorithm and a recoil-sensitive axis: This was recently studied in ref.~\cite{Bain:2016rrv}. The TMD fragmentation function involves rapidity divergences. One can define a factor consisting of (collinear-)soft modes which cancels these rapidity divergences. However, this (collinear-)soft radiation will displace the jet axis and contaminate the transverse momentum distribution, again introducing a sensitivity to NGLs.\footnote{From the direct two-loop calculations for related jet shapes in the soft approximation~\cite{Kelley:2011aa}, one sees that NGLs are present for all jet radii. Unlike in ref.~\cite{Kelley:2011aa}, the out-of-jet radiation is not restricted here, making it effectively equal to the partonic center-of-mass energy.}.
\item Inclusive production with a jet algorithm and a recoil-free axis: the observable is purely collinear, making it universal and free of NGLs. This is the case we focus on.

\end{itemize}

We now briefly outline our framework: For definiteness we focus on hadronic collisions with energetic jets that are not particularly close to each other or the beam axis (i.e.~central jets). Our approach is easily extendable using e.g.~refs.~\cite{Bauer:2011uc,Larkoski:2015zka,Pietrulewicz:2016nwo}. Consider the measurement of the longitudinal momentum fraction $z_h$ and transverse momentum $z_h \vecb k$ (with respect to the jet axis) of an energetic hadron $h$ inside a jet.\footnote{This definition of  $\vecb k$ ensures that it is a partonic observable and thus perturbatively calculable.}
To leading approximation, the soft radiation outside the jet cannot affect the production of a hadron inside a jet. However, as illustrated in \fig{recoil}, the measurement of the transverse momentum of a hadron with respect to the standard jet axis is sensitive to the soft radiation inside the jet. This is not the case when using a recoil-insensitive axis, which is determined by the configuration of the energetic (collinear) radiation. Under the assumption that the jet radius $R \ll 1$ (the case $R\sim 1$ will also be discussed), collinear factorization leads to
%%% 
\begin{align}\label{eq:match_si1}
  \frac{\df \si_h}{\df p_T\, \df \eta\, \df^2 \vecb k\, \df z_h}
  = \sum_i \int\! \frac{\df x}{x}\, \hat \si_i\Big(\frac{p_T}{x}, \eta, \mu\Big)\, \cG_{i\to h}(x, p_T R, \vecb k, z_h,\mu) \big[1 + \ord{R^2}\big]
\,.\end{align}
%%%
The partonic cross section $\hat \si$ encodes the hard scattering producing the parton $i$ with transverse momentum $p_T/x$ and rapidity $\eta$, with respect to the beam axis. The fragmenting jet function $\cG$ describes the fraction $x$ of the parton energy that goes into the jet, as well as the fragmentation of the hadron inside the jet with momentum fraction $z_h$ and transverse momentum $z_h \vecb k$. The function $\cG$ obeys a collinear renormalization group equation. A further factorization of this cross section can be achieved when $p_T R\gg |\vecb k|$ and/or $|\vecb k|\gg\Lambda_{QCD}$, which is discussed in detail in \sec{fact}. In particular, for $p_T R\gg |\vecb k|$ we can separate the effect of the jet boundary $B$ from the fragmentation, leading to a new JTMD fragmentation function $D_{k\to h}$,
%%%
\begin{align}\label{eq:match_G2}
\cG_{i\to h}(x, p_T R, \vecb k, z_h,\mu) = \sum_k \int\! \frac{\df y}{y}\, B_{ik}(x, p_T R, y,\mu)\, D_{k \to h}\Big(\vecb k, \frac{z_h}{y},\mu\Big) \bigg[1 + \ORd{\frac{\vecb k^2}{p_T^2 R^2}} \bigg]
\,.\end{align}
%%%   
Since $D_{k\to h}$ is a purely collinear object, it is automatically universal, i.e.~insensitive to the type of process or number of jets. It also does not involve rapidity divergences, unlike the classical TMD fragmentation functions.

The paper is organized as follows:
We start by outlining the differences between the classical TMDs and the JTMDs that are considered in this work in \sec{framework}. We also define all ingredients that enter in our fractorization theorems and discuss their renormalization. A discussion of recoil-free jet definitions in the context of a simple example in given in \app{recoil}, including a one-loop calculation.  The winner-take-all recombination scheme~\cite{Salam:WTAUnpublished,Bertolini:2013iqa} that we use to obtain a recoil-free jet axis is summarized in \app{cluster}.
 In \sec{fact}, we show how \eq{match_si1} can be further factorized, depending on the hierarchy between $p_T R$, $|\vecb k|$ and $\Lambda_{QCD}$. We also treat the case when $R$ is not small. We have calculated the one-loop matching coefficients and present these in \sec{matching}. In \sec{res} some first numerical results based on a moment analysis are presented. We conclude in \sec{concl}, discussing the wide range of potential applications of our framework.

%%%%%%%%%%%%%%%%%%%%%%%%%%%%%%%%%%%%%%%%%%%%%%%%%%%%%%%%%%%%%%%%%%%%%%%%%%%%%%%%
\section{Framework}
\label{sec:framework}
%%%%%%%%%%%%%%%%%%%%%%%%%%%%%%%%%%%%%%%%%%%%%%%%%%%%%%%%%%%%%%%%%%%%%%%%%%%%%%%%

We use light-like vectors $n$ and $\bn$ with $n\cdot\bn=2$ to introduce the light-cone coordinates used here
%%%
\begin{align}\label{eq:LC_Coor}
  v^\mu = v^- \frac{n^\mu}{2} + v^+\, \frac{\bn^\mu}{2} + v_\perp^\mu
  \,, \quad
  v^- = \bn \sdt v
  \,, \quad
  v^+ = n \sdt v
\,.\end{align}
%%%
The time-like and space-like component of a vector are indicated  by $(v_0,\vec v)$, so that 
%%%
\begin{align}
v^2=v_0^2-\vec v^{\,2}=v^+v^-+v_\perp^2=v^+v^--\vecb v^2
\,.\end{align}
%%%
In the language of soft-collinear effective theory (SCET)~\cite{Bauer:2000ew, Bauer:2000yr, Bauer:2001ct, Bauer:2001yt}, if we assign a power counting for the collinear momenta as
%%%
\begin{align}
p_n=(\nbar\cdot p_n,n\cdot p_n, {\mathbf p_n})\sim Q(1,\lambda^2,\lambda),
\,,\end{align} 
%%%
where the power counting parameter $\la \ll 1$ is set by the specific measurement,
then we can define a power counting for the soft radiation as
%%%
\begin{align} \label{eq:soft}
p_{s}&\sim Q(\lambda^{\beta},\lambda^{\beta},\lambda^{\beta})
\,.\end{align}
%%%
For $\beta=1$ this is referred to as soft radiation, and for $\beta=2$ it is called ultra-soft.

%===============================================================================
\subsection{Standard transverse momentum dependent fragmentation functions}
%===============================================================================

Before  arriving at the formulation of the JTMDs, it is instructive to recall some properties of the classical unpolarized TMD fragmentation functions~\cite{Collins:2011zzd, Echevarria:2016scs},
%%%
\begin{eqnarray} \label{eq:def_FF_opsand}
\nn
\Delta_{q\to h}(z_h,\vecb b_T)&=&\frac{1}{4 z_h N_c}\sum_X\int\! \frac{\df\xi^+}{4\pi}\,e^{-ip^-_h\xi^+/(2 z_h)}\,
\langle 0|T\!\[\tilde W_{Tn}^{\dagger}q_j\]_{a}\!\!\Big(\frac{\xi}{2}\Big)
|X,h\rangle \gamma^-_{ij} \langle X,h|\bar T\!\[\bar q_i \,\tilde W_{Tn}\]_{a}\!\!\Big(-\frac{\xi}{2}\Big)|0\rangle,
\\ 
\Delta_{g\to h}(z_h,\vecb b_T)&=&\frac{-1}{2(1-\epsilon)p_h^-(N_c^2-1)}\sum_X\int\! \frac{\df\xi^+}{4\pi}\,e^{-ip^-_h\xi^+/(2 z_h)}
\nn \\ &&
\times\langle 0|T\!\[\tilde W_{Tn}^{\dagger}F^{-\mu}\]_{a}\!\!\Big(\frac{\xi}{2}\Big)
|X,h\rangle g_{\mu\nu} \langle X,h|\bar T\!\[F^{-\nu} \,\tilde W_{Tn}\]_{a}\!\!\Big(-\frac{\xi}{2}\Big)|0\rangle,
\end{eqnarray}
%%%
where $\xi=(\xi^+,0^-,\vecb b_T)$. The variable conjugate to the impact parameter $\vecb b_T$ is $\vecb k$, which 
is the transverse momentum of the hadron divided by its momentum fraction.  The sum runs over all intermediate states $X$, and $X$ does not include the hadron $h$. 
 The Wilson lines $\tilde W_{Tn}(x)$ depend on  the coordinate $x$ and continue to the light-cone infinity along the vector $n$, where it is connected by a transverse link to the transverse infinity (as indicated by the subscript $T$)~\cite{Idilbi:2010im,GarciaEchevarria:2011md}.
 The representations of the SU(3) generators inside the Wilson lines correspond to that of the parton (fundamental for quark, adjoint for gluon), and repeated color indices are summed over.

It is important to emphasize that implicit in these definitions of the TMDFFs a specific axis choice has been made, namely that the $n$ direction is along the hadron $h$. By performing a change of coordinates (or reparametrization~\cite{Manohar:2002fd}), it follows that this corresponds to measuring the transverse momentum of the hadron with respect to the axis lying along the total momentum of all particles in the intermediate state. For fragmentation in $e^+e^-\to$ hadrons, this axis is equivalent to the thrust axis.

These TMDs appear in processes like SIDIS or $e^+ e^- \to $ hadrons and involve both ultraviolet (UV) and rapidity divergences that require renormalization through a soft factor. Consequently the renormalization group equations obeyed by these TMDs involve a resummation of both the UV and rapidity factorization scales. The factorization of these processes is often described in impact parameter space and the hadrons in the final state must in principle be detected on the whole phase space. In the limit of large transverse momentum $\vecb k$, or equivalently $\vecb b_T\to 0$, the TMDFFs can be matched onto the standard (integrated) fragmentation functions. These are defined as~\cite{Collins:1981uw} 
%%%
\begin{align}\nn
d_{q\to h}(z_h)&=\frac{1}{4 z_h N_c}\sum_X\int\! \frac{\df\xi^+}{4\pi}\,e^{-ip^-_h\xi^+/(2 z_h)}\,
%\\\nn &\times
 \langle 0|T\!\[\tilde W_{Tn}^{\dagger}q_j\]_{a}\!\!\Big(\frac{\xi^+}{2}\Big)
|X,h\rangle \gamma^-_{ij} \langle X,h|\bar T\!\[\bar q_i \,\tilde W_{Tn}\]_{a}\!\!\Big(-\frac{\xi^+}{2}\Big)|0\rangle,
\\
d_{g\to h}(z_h)&=\frac{-1}{2(1-\epsilon)p_h^-(N_c^2-1)} \sum_X\int\! \frac{\df\xi^+}{4\pi}\,e^{-ip^-_h\xi^+/(2 z_h)}
 \nn\\
 & \quad \times \langle 0|T\!\[\tilde W_{Tn}^{\dagger}F^{-\mu}\]_{a}\!\!\Big(\frac{\xi^+}{2}\Big)
\sum_X|X,h\rangle g_{\mu\nu}  \langle X,h|\bar T\!\[F^{-\nu} \,\tilde W_{Tn}\]_{a}\!\!\Big(-\frac{\xi^+}{2}\Big)|0\rangle.
\label{eq:ff}
\end{align}
%%%

%===============================================================================
\subsection{Definitions for TMD fragmentation inside a jet}
%===============================================================================

We now turn to the operator definitions of the JTMDFFs. The key observation for defining a recoil-free observable, which mitigates its soft sensitivity, is that the recoil of soft radiation translates the whole of the collinear sector coherently in the transverse momentum plane. Therefore, if we define a jet axis that also recoils coherently with the soft radiation, any collinear measurement relative to that axis will be insensitive to these recoil effects. The simplest definition of a recoil-free axis is given via recombination jet algorithms, as summarized in \app{cluster}. The basic logic is that given a list of particles, we have a measure to decide what members of the list should be grouped together as if they came from a single hard progenitor. At each stage of the recombination two particles are merged, and we must decide what the direction is of the ``particle" formed by the merged particles. In the winner-take-all (WTA) scheme, this is chosen to be the direction of the more energetic of the two daughters~\cite{Salam:WTAUnpublished,Bertolini:2013iqa}. This scheme is inherently recoil free, since the winners of the axis direction are always the most energetic clusters of particles in the jet.

Having a recoil-free axis in a recombination algorithm is then simply a matter of the merger step. Thus any specific recombination algorithm can be made recoil free, and satisfies \eq{match_si1}. However, whether one can further factorize collinear splittings landing near the boundary of the jet and those deep inside, depends on the specific measure used to decide which particles will be merged. We will argue in \sec{fact} that this is the case for the Cambridge/Aachen \cite{Dokshitzer:1997in,Wobisch:1998wt,Wobisch:2000dk} and anti-$k_T$ measures \cite{Cacciari:2008gp}, provided the transverse momentum is sufficiently small such that the hadron is not at the edge of the jet. 

In what follows, we call the light-cone directions $n, \nbar$ introduced in \eq{LC_Coor} the \emph{fiducial} light-cone directions. These are not dynamical, and are simply necessary to define the collinear sector and its gauge-invariant operators. 
The price paid for a recoil-free axis is that the axis is sensitive to the precise final state configuration of the collinear emissions relative to each other. This is not the case for a thrust axis, which is essentially a conserved quantity under the collinear splittings, and thus independent of the dynamics\footnote{Indeed, from a factorization point of view, this is what makes the thrust axis natural. The light-cone directions used to define the collinear sector should not depend on the specific configuration of collinear particles, since the factorization itself is unphysical (e.g.~it depends on a specific renormalization point). However, the only \emph{physical} jet axis that is independent of the collinear final state is the direction of total momentum flow, since it is conserved.}. We can demand that the jet has zero transverse momentum with respect to the fiducial light-cone directions, and if we gave it a non-zero transverse momentum with respect to these directions, we would find that it could be translated away in the course of the calculation. A one-loop example of this phenomena is given in \app{recoil}. This captures the notion that the definition of the collinear sector is arbitrary up to translations satisfying a particular power counting, known in the effective theory literature as reparametrization invariance \cite{Luke:1992cs, Manohar:2002fd}.
Ultimately, it is the measurements imposed on the collinear sector that determine the power counting of the allowed reparametrization:
 for recoil-sensitive measurements, the reparametrizations are restricted to those satisfying an ultra-soft power counting~\cite{Bauer:2008dt}. However, for recoil-insensitive measurements, reparametrizations with a soft scaling (see \eq{soft}) are allowed.

We now present the QCD matrix elements for our fragmenting jet functions and JTMD  fragmentation functions. 
The momentum fraction is defined as
%%%
\begin{align}
z_h=\frac{p_h^-}{p_J^-},
\end{align} 
%%%
where $p_h^-$ and $p_J^-$ are the large momentum component of the hadron and jet, respectively. Then we write:
%%%
\begin{align} \label{eq:Gdef}
\cG_{q\to h}(x,p_T R,\vecb k,z_h)&=\frac{1}{4 x N_c}\sum_{X} \sum_{J/h}
 \int \frac{\df\xi^+}{4\pi}\,
e^{-i p_J^- \xi^+/(2 x)}  \de\Big(z_h - \frac{p_h^-}{p_J^-}\Big)
\int\! \df k_A\, \de^{(3)}\Big(\vec{k}- \frac{\vec p_h}{z_h}\Big)
\\ & \quad
\times\langle 0|T\[\tilde W_{Tn}^{\dagger}q_j\]_{a}\Big(\frac{\xi^+}{2}\Big)
|X,h\in J\rangle\, \gamma^-_{ij} \, \langle X,h\in J|\bar T\[\bar q_i \,\tilde W_{Tn}\]_{a}\Big(-\frac{\xi^+}{2}\Big)|0\rangle,
\nn \\ 
\cG_{g\to h}(x,p_T R,\vecb k,z_h)&=\frac{-1}{2(1-\epsilon)p_J^-(N_c^2-1)}
\sum_{X} \sum_{J/h}
 \int \frac{\df\xi^+}{4\pi}\,
e^{-ip_J^-\xi^+/(2 x)}  \de\Big(z_h - \frac{p_h^-}{p_J^-}\Big)
\int\! \df k_A\, \de^{(3)}\Big(\vec{k}- \frac{\vec p_h}{z_h}\Big)
\nn \\ & \quad
\times\langle 0|T\[\tilde W_{Tn}^{\dagger}F^{-\mu}\]_{a}\Big(\frac{\xi^+}{2}\Big)
|X,h\in J\rangle g_{\mu\nu} \langle X, h\in J|\bar T\[F^{-\nu} \,\tilde W_{Tn}\]_{a}\Big(-\frac{\xi^+}{2}\Big)|0\rangle,
\nn \end{align}
%%%
Here, the sum runs over the jets $J$ in the final state, with momentum $p_J$. The hadron $h$ is part of $J$, but its phase-space integral is not included in the sum on $J$, as indicated by $J/h$. The unit vector $\vec A_J$ along the jet axis is obtained in the WTA scheme, as discussed above and in \app{cluster}. In \eq{Gdef} the integration over $k_A = A_J \cdot p_h$, the component of the momentum $\vec k$ along the axis $\vec A_J$, ensures that $\vecb k$ picks up the components transverse to this axis. These fragmenting jet functions are a more differential version of the (semi-inclusive) fragmenting jet function~\cite{Procura:2009vm,Procura:2011aq,Kang:2016ehg,Dai:2016hzf}, see also \sec{sumrule}.

When $ p_T R\gg |\vecb k|$ we can perturbatively match the  functions $\cG_{i\to h}(x,p_T R,\vecb k,z_h)$ onto the JTMDFFs $D_{j\to h}(\vecb k,z_h)$, which are  defined as
%%%
\begin{align}\label{eq:Ddef}
D_{q\to h}(\vecb k,z_h)&=\frac{1}{4 z_h N_c}\sum_{X} 
 \int \frac{\df\xi^+}{4\pi}\,  e^{-i p_h^- \xi^+/(2 z_h)}  
\int\! \df k_A\, \de^{(3)}\Big(\vec{k}- \frac{\vec p_h}{z_h}\Big)
\nn \\ & \quad
\times\langle 0|T\[\tilde W_{Tn}^{\dagger}q_j\]_{a}\Big(\frac{\xi^+}{2}\Big)
|X,h\rangle \gamma^-_{ij} \langle X,h|\bar T\[\bar q_i \,\tilde W_{Tn}\]_{a}\Big(-\frac{\xi^+}{2}\Big)|0\rangle,
\nn \\
D_{g\to h}(\vecb k,z_h)&=\frac{-1}{2(1-\epsilon)p_h^-(N_c^2-1)}
\sum_X
 \int \frac{\df\xi^+}{4\pi}\,  e^{-ip_h^-\xi/(2 z_h)}  
\int\! \df k_A\, \de^{(3)}\Big(\vec{k}- \frac{\vec p_h}{ z_h}\Big)
\nn \\ & \quad
\times\langle 0|T\[\tilde W_{Tn}^{\dagger}F^{-\mu}\]_{a}\Big(\frac{\xi^+}{2}\Big)
|X,h\rangle g_{\mu\nu} \langle X, h|\bar T\[F^{-\nu} \,\tilde W_{Tn}\]_{a}\Big(-\frac{\xi^+}{2}\Big)|0\rangle,
\end{align}
%%%
In this expression the boundary of the jet has been expanded to infinity, so $X$ runs over an unrestricted set of states that is independent of the jet definition, and $h$ is not part of $X$. The only dependence on the jet algorithm is through the definition of the jet axis. Note that the only difference with \eq{def_FF_opsand} is the axis with respect to which the transverse momentum is measured.

%===============================================================================
\subsection{Renormalization}
\label{sec:renorm}
%===============================================================================

The renormalized fragmentation functions are defined through~\cite{Collins:1981uw}
%%%
\begin{align} \label{eq:D_ren}
  d_{i\to h}^{\text{bare}}(z_h) 
  & = \sum_j \int \frac{\df z}{z}\, Z_{ij}\Big(\frac{z_h}{z},\mu\Big)\, d_{j\to h}(z,\mu)
  \,,
\end{align}
%%%
leading to the following renormalization group equation (RGE) 
%%%
\begin{align} \label{eq:RGE_D}
  \mu \frac{\df} {\df\mu} d_{i\to h}(z_h,\mu) 
  &= \sum_j \int \frac{\df z}{z}\, \ga_{ij}\Big(\frac{z_h}{z}, \mu\Big) d_{j\to h}(z, \mu)
  \,, \nn \\
  \gamma_{ij}(z_h,\mu) &= -\! \int \frac{\df z}{z}\, Z^{-1}_{ik}\Big(\frac{z_h}{z},\mu\Big) \,
  \mu\frac{\df}{\df\mu} Z_{kj}(z,\mu) 
  \,.
\end{align}
%%%

The fragmenting jet function $\cG$ has the same renormalization and thus RGE as the fragmentation function, but in the $x$ variable~\cite{Kang:2016mcy, Dai:2016hzf, Kang:2016ehg}
%%%
\begin{align} \label{eq:RGE_G}
\cG_{i\to h}^{\text{bare}}(x, p_T R, \vecb k, z_h,\mu) = \sum_j \int \frac{\df x'}{x'}\, Z_{ij}\Big(\frac{x}{x'},\mu\Big)\, \cG_{j\to h}(x', p_T R, \vecb k, z_h,\mu) 
\,.\end{align}
%%%
The RGE of the matching coefficients $\cJ$ in \eq{match_G} follows from inserting \eqs{RGE_D}{RGE_G} in \eq{match_G}, and thus involves a DGLAP evolution in both $x$ and $z$.

The renormalization of the JTMD fragmentation function has the same structure as that of the standard fragmentation function, 
%%%
\begin{align} \label{eq:RGE_D_k}
  D_{i\to h}^{\text{bare}}(\vecb k, z) 
  & = \sum_j \int \frac{\df z'}{z'}\, Z_{ij}'\Big(\frac{z}{z'},\mu\Big)\, D_{j\to h}(\vecb k, z',\mu)
  \,,
\end{align}
%%%
however it involves a different renormalization factor, $Z' \neq Z$. The RGE thus has the same structure as \eq{RGE_D} but the anomalous dimension is modified to $\ga'$. 

The all-orders anomalous dimensions are given by 
%%%
\begin{align}
  \ga_{ij}(z,\mu) &= P_{ji}(z,\mu)
  \,,\nn \\ 
  \ga_{ij}'(z,\mu) &= \theta\Big(z \geq \frac12\Big)\, P_{ji}(z,\mu)
\,,\end{align}
%%%
where $P$ denote the DGLAP splitting functions~\cite{Gribov:1972ri,Altarelli:1977zs,Dokshitzer:1977sg}. At one-loop order this follows directly from our calculation. In \app{anom}, we argue this relationship is true to all orders, and the corresponding expressions in moment space are given to one loop.

%===============================================================================
\subsection{Sum rule}
\label{sec:sumrule}
%===============================================================================

The jet definition restricts the maximum transverse momentum $|\vecb k|$ of the hadron. The transverse momentum $\vecb k$ may therefore safely be integrated over
%%%
\begin{align}
\int \! \df^2 \vecb k\,\cG_{i\to h}(x, p_T R, \vecb k, z_h,\mu) = \cG_{i\to h}(x, p_T R, z_h,\mu)
\,,\end{align}
%%%
to yield the (semi-inclusive) fragmenting jet function~\cite{Kang:2016ehg,Dai:2016hzf}. The same is not true for the TMD fragmentation function, which has a different renormalization than the fragmentation function.

%%%%%%%%%%%%%%%%%%%%%%%%%%%%%%%%%%%%%%%%%%%%%%%%%%%%%%%%%%%%%%%%%%%%%%%%%%%%%%%%
\section{Jet factorization and TMD fragmentation}
\label{sec:fact}
%%%%%%%%%%%%%%%%%%%%%%%%%%%%%%%%%%%%%%%%%%%%%%%%%%%%%%%%%%%%%%%%%%%%%%%%%%%%%%%%

Our starting point is the cross section for producing a jet with transverse momentum $p_T$ and rapidity $\eta$, containing a hadron with momentum fraction $z_h$ and transverse momentum $z_h \vecb k$,
%%% 
\begin{align}\label{eq:match_si2}
  \frac{\df \si_h}{\df p_T\, \df \eta\, \df^2 \vecb k\, \df z_h}
  = \sum_i \int\! \frac{\df x}{x}\, \hat \si_i\Big(\frac{p_T}{x}, \eta, \mu\Big)\, \cG_{i\to h}(x, p_T R, \vecb k, z_h) \big[1 + \ord{R^2}\big]
\,.\end{align}
%%%
This observable is insensitive to soft radiation, since the transverse momentum $\vecb k$ is measured relative to a recoil-insensitive axis. The above equation thus follows from collinear factorization for $R \ll 1$. The partonic cross section $\hat \si$ encodes the hard scattering that produces the parton $i$ with transverse momentum $p_T/x$ and rapidity $\eta$, with respect to the beam axis. The fragmenting jet function $\cG$ was defined in \eq{Gdef} and describes the longitudinal momentum fraction $x$ of the parton that goes into the jet, as well as the fragmentation of the hadron inside the jet.
Depending on the relative hierarchy between $p_T R$, $|\vecb k|$ and $\lqcd$, \eq{match_si2} admits a further factorization.

%===============================================================================
\subsection{Factorization of fragmentation from perturbative radiation}
%==============================================================================

If $p_T R \sim |\vecb k | \gg \lqcd$, the perturbative dynamics that resolves the jet boundary and generates the transverse momentum factorizes from the nonperturbative fragmentation~\cite{Procura:2009vm,Jain:2011xz},
%%%
\begin{align}\label{eq:match_G}
\cG_{i\to h}(x, p_T R, \vecb k, z_h,\mu) = \sum_j \int\! \frac{\df z}{z}\, \cJ_{ij}\Big(x, p_T R,\vecb k, \frac{z_h}{z},\mu\Big) d_{j\to h}(z,\mu) \bigg[1 + \ORd{\frac{\lqcd^2}{\vecb k^2}} \bigg]
\,.\end{align}
%%%
The matching coefficient $\cJ_{ij}$ describes the formation of a jet with momentum fraction $x$ of the initial parton $i$, containing a parton $j$ with momentum fraction $z_h/z$ and transverse momentum $\vecb k$. The (standard) fragmentation function $d_{j \to h}$ describes how this parton $j$ produces a hadron moving in the same direction with a momentum fraction $z_h/z \times z = z_h$, see \eq{ff}.

%===============================================================================
\subsection{Factorization of TMD fragmentation from jet definition}
%==============================================================================

For $p_T R \gg | \vecb k | \gg \lqcd$, a judicious choice of jet axis enables one to separate the effect of the jet boundary and the generation of the perturbative transverse momentum of the hadron,
%%%
\begin{align}\label{eq:match_J}
  \cJ_{ij}(x, p_T R, \vecb k, z,\mu) = \sum_k \int\! \frac{\df y}{y}\, 
    B_{ik}(x, p_T R, y,\mu)\, C_{kj}\Big(\vecb k, \frac{z}{y},\mu\Big) \bigg[1 + \ORd{\frac{\vecb k^2}{p_T^2 R^2}} \bigg]
\,,\end{align}
%%%
due to a second collinear factorization at angular scales $r = |\vecb k|/p_T \ll R$. This requires the factorization of the amplitude and the measurement, which we discuss in turn. 

For the amplitude to factorize, there must be an energetic parton within an angular distance $r$ of the axis. This is ensured for the winner-take-all axis, which by construction is always along the direction of such a parton. The hadron will fragment from this parton in order to be enhanced in the small $|\vecb k|$ limit. Of course there can be additional partons in the vicinity of the axis. If they are produced as splittings from an initial parton, their effect is captured by $C$ in \eq{match_J}. The case where independent emissions at angular scales $R$ randomly happen to be within a distance $r$ is power suppressed by $r/R$.

\begin{figure}
\centering
\includegraphics[width=0.55\textwidth]{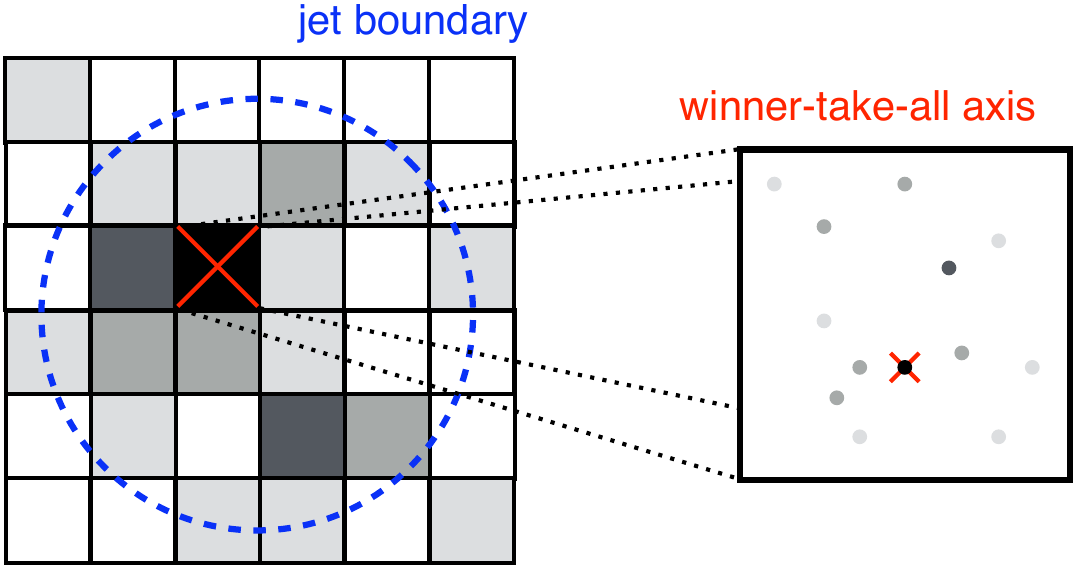}%
\caption{Factorization of the axis finding between the angular scale $r =  |\vecb k|/p_T$ and $R$, with $r \ll R$.
\label{fig:wta_fact}}
\end{figure}

For the measurement to factorize as in \eq{match_J}, the axis finding must be ``recursively local". What we mean is that the jet axis can be determined within a angular distance of $r \ll R$ by only considering collinear emissions at angular scales of order $R$, whereas a more precise determination of the axis position only requires knowledge of radiation within an angular distance $r$. A more concrete way of thinking about this is illustrated in \fig{wta_fact}: we ``pixelate" the measurement into regions of angular size $r$, and the total energy of each pixel is sufficient to determine the pixel containing the axis. The position of the axis within the pixel only relies on the energy distribution within an angular size $r$. Collinear splittings inside the pixel only shift the axis an amount of order $r$ and are thus power suppressed by order $r/R$ for radiation at the jet boundary. This guarantees the simple convolution structure in \eq{match_J}, where the collinear radiation at angular scales $R$ and $r$ only communicate through a single variable: the energy fraction of the ``pixel" containing the winner-take-all axis. 

\begin{figure}
\centering
\includegraphics[width=0.4\textwidth]{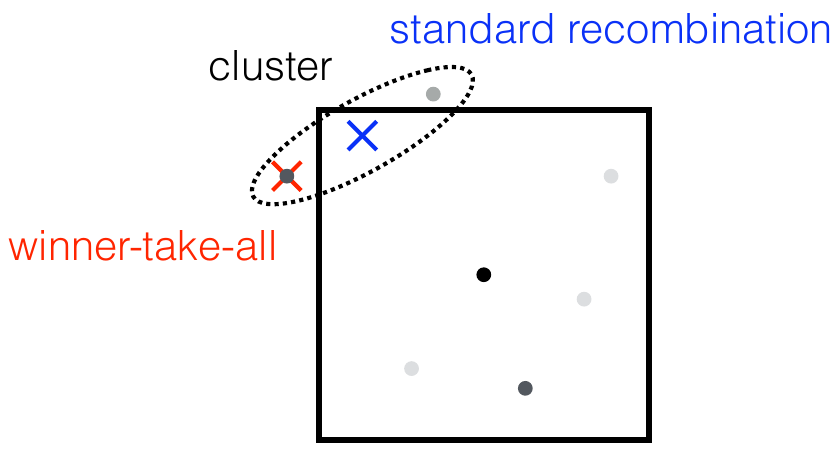}%
\caption{The standard recombination scheme allows particles outside a pixel to be clustered into a pixel (blue), \emph{without} being clustered with the pixel. This is not the case for the winner-take-all scheme (red).
\label{fig:cluster_fail}}
\end{figure}

When we argue for this recursively local picture of the axis determination, we must establish two properties: radiation within the pixel that eventually contains the jet axis will be preferentially clustered together first and the configuration of the radiation outside of this pixel does not interfere with the constituents of the pixel, except perhaps at the boundary. The Cambridge/Aachen clustering algorithm~\cite{Dokshitzer:1997in,Wobisch:1998wt} with the winner-take-all recombination scheme naturally has these properties, since it is solely based on angular distances. By definition, most of the radiation within the pixel is at a closer angular distance to each other than to radiation outside the pixel, and this will be recombined first, except for possible splittings at the boundary. Radiation far from the pixel, e.g.~at the jet boundary, will not be clustered in too early. Radiation outside the pixel that is clustered together will not interfere with the clustering history inside, since the winner-take-all axis always lies on a particle at each step in the recombination. Specifically, two particles outside of a pixel can never be recombined to give a ``shadow'' particle within the pixel, as illustrated in \fig{cluster_fail}, regardless of the ordering in which particles get recombined.

The key difference between anti-$k_T$ and Cambridge/Aachen is the order in which radiation is clustered. As is well established, anti-$k_T$ clusters the most energetic radiation first. By definition, the pixel which will contain the winner-take-all axis in anti-$k_T$ will be clustered preferentially, since this is the most energetic region and is where the algorithm will start to cluster. However, radiation around this pixel may not first be clustered with each other but could directly be clustered with that pixel. Nevertheless, the collinear splittings inside the pixel containing the winner-take-all axis still factorize from the splittings at the jet boundary, i.e.~changes in the angle of the jet axis due to collinear splittings inside the pixel will be of order $r$, and the effect at the jet boundary is thus power suppressed by $r/R$.

Thus we have shown that with the winner-take-all axis, the Cambridge/Aachen and the anti-$k_T$ algorithms satisfy the factorization in \eq{match_J}. Note the importance of establishing these all-orders properties, since the one-loop calculation of the matching coefficients in this paper only involve final states with at most two partons, in which case the winner-take-all axis is simply along the most energetic parton.

For $p_T R \gg |\vecb k| \sim \lqcd$, we can separate the effect of the jet boundary from the fragmentation, but cannot calculate the nonperturbative transverse momentum,
%%%
\begin{align}\label{eq:match_G2}
\cG_{i\to h}(x, p_T R, \vecb k, z_h,\mu) = \sum_k \int\! \frac{\df y}{y}\, B_{ik}(x, p_T R, y,\mu)\, D_{k\to h}\Big(\vecb k, \frac{z_h}{y},\mu\Big) \bigg[1 + \ORd{\frac{\vecb k^2}{p_T^2 R^2}} \bigg]
\,.\end{align}
%%%
The JTMD fragmentation function $D$ that arises here is the universal object anticipated before, and is defined through \eq{Ddef}. As we may also obtain \eq{match_J} by a further factorization of \eq{match_G2} for $|\vecb k| \gg \lqcd$, consistency implies that the same boundary function $B$ enters in these equations and
%%%
\begin{align}\label{eq:match_D}
D_{k\to h}(\vecb k, z_h,\mu) = \sum_j \int\! \frac{\df z}{z}\, C_{kj}\Big(\vecb k, \frac{z_h}{z},\mu\Big)\, d_{j\to h}(z,\mu) \bigg[1 + \ORd{\frac{\lqcd^2}{\vecb k^2}} \bigg]
\,.\end{align}
%%%

%===============================================================================
\subsection{Factorization for large radius jets}
%==============================================================================

So far we have always assumed that the jet radius $R$ is small, allowing for the factorization in \eq{match_si2}. However, when $R$ is large the jet at scale $p_T R$ cannot be factorized from the hard scattering at scale $p_T$. In this case we can still factorize the JTMD fragmentation functions when $\vecb k^2 \ll p_T^2$,
%%% 
\begin{align}\label{eq:match_si3}
  \frac{\df \si_h}{\df p_T\, \df \eta\, \df^2 \vecb k\, \df z_h}
  = \sum_k \int\! \frac{\df y}{y}\, \bar \si_k(p_T, \eta, R, y, \mu)\, D_{k \to h}\Big(\vecb k, \frac{z_h}{y},\mu\Big) \bigg[1 + \ORd{\frac{\vecb k^2}{p_T^2}}\bigg]
\,.\end{align}
%%%
The partonic cross section $\bar \si$ now describes the hard scattering $\hat \si_i$ and the jet boundary effects $B$. Indeed, in the limit $R \ll 1$, 
%%% 
\begin{align}
  \bar \si_k(p_T, \eta, R, y, \mu) = 
  \sum_i \int\! \frac{\df x}{x}\, \hat \si_i\Big(\frac{p_T}{x}, \eta, \mu\Big)\, B_{ik}(x, p_T R, y,\mu) \big[1 + \ord{R^2}\big]
\,.\end{align}
%%%

%%%%%%%%%%%%%%%%%%%%%%%%%%%%%%%%%%%%%%%%%%%%%%%%%%%%%%%%%%%%%%%%%%%%%%%%%%%%%%%%
\section{NLO maching coefficients}
\label{sec:matching}
%%%%%%%%%%%%%%%%%%%%%%%%%%%%%%%%%%%%%%%%%%%%%%%%%%%%%%%%%%%%%%%%%%%%%%%%%%%%%%%%

In this section we summarize the one-loop matching coefficients that appear in \sec{fact}.

%===============================================================================
\subsection{Fragmenting jet function}
%===============================================================================

The matching coefficients that enter in \eq{match_G} are given by 
%%%
\begin{align} 
&\cJ_{ij}^\zero(x, p_T R, \vecb k, z,\mu) 
= \de_{ij}\, \de^2(\vecb k) \de(1-x)\de(1-z)
\,, \\
%%%%%%%%%%%%%%%%%%
&\cJ_{qq}^\one(x, p_T R, \vecb k, z,\mu) 
\nn \\ & \quad
= 
\frac{\al_s C_F}{2\pi}\,
\bigg( \frac{1}{\pi}\,
 \frac{1}{\mu^2}\,\frac{1}{(\vecb k^2/\mu^2)}_+\!\! \de(1-x)\,\theta\Big(\frac12 \geq z\Big)\theta\big(p_T R  \geq |\vecb k|\big)\, \frac{1+z^2}{1-z}
\nn \\ & \qquad
- \de^2(\vecb k)\de(1-z)\theta(1-x) \bigg\{ (1+x^2)\Big[\Big(\frac{1}{1-x}\Big)_+\ln \Big(\frac{p_T^2 R^2}{\mu^2}\Big)+2\Big(\frac{\ln (1-x)}{1-x}\Big)_+ \Big] + 1-x \bigg\} 
\nn \\ & \qquad
+ \de^2(\vecb k)\de(1-x)\bigg\{\theta\Big(1\geq z\geq \frac12\Big)\bigg[ (1+z^2)\Big[\Big(\frac{1}{1-z}\Big)_+\ln \Big(\frac{p_T^2 R^2z^2}{\mu^2}\Big)+2\Big(\frac{\ln (1-z)}{1-z}\Big)_+ \Big]  + 1-z \bigg]
\nn \\ & \qquad
+\theta\Big(\frac12\geq z\Big)\Big[2\frac{1+z^2}{1-z}\ln\big(z(1-z)\big)+(1-z)\Big]\bigg\} \bigg)
\,, \\
%%%%%%%%%%%%%%%%%%
&\cJ_{qg}^\one(x, p_T R, \vecb k, z,\mu) 
\nn \\ & \quad
= \frac{\al_s C_F}{2\pi}\,
\bigg(\frac{1}{\pi}\, \frac{1}{\mu^2}\,\frac{1}{(\vecb k^2/\mu^2)}_+\!\! \de(1-x)\,\theta\Big(\frac12 \geq z\Big)\theta\big(p_T R  \geq |\vecb k|\big)\, \frac{1+(1-z)^2}{z}
\nn \\ & \qquad
- \de^2(\vecb k)\de(1-z)\theta(1-x) \bigg\{ \frac{1+(1-x)^2}{x}\ln \Big(\frac{p_T^2 R^2(1-x)^2}{\mu^2}\Big) + x\bigg\} 
\nn \\ & \qquad
+ \de^2(\vecb k)\de(1-x)\bigg\{\theta\Big(1\geq z\geq \frac12\Big)\Big[ \frac{1+(1-z)^2}{z}\ln\Big( \frac{p_T^2 R^2z^2(1-z)^2}{\mu^2}\Big) +z\Big]
\nn \\ & \qquad
+\theta\Big(\frac12\geq z\Big)\Big[2\frac{1+(1-z)^2}{z}\ln\big(z(1-z)\big)+z\Big]\bigg\} \bigg)
\,, \\
%%%%%%%%%%%%%%%%%%
&\cJ_{gg}^\one(x, p_T R, \vecb k, z,\mu) 
\nn \\ & \quad
= 
\frac{\al_s C_A}{2\pi}\,
\bigg(\frac{1}{\pi}\,\frac{1}{\mu^2}\,\frac{1}{(\vecb k^2/\mu^2)}_+\!\! \de(1-x)\,\theta\Big(\frac12 \geq z\Big)\theta\big(p_T R  \geq |\vecb k|\big)\,\frac{2(1-z+z^2)^2}{z(1-z)}
\nn \\ & \qquad
- \de^2(\vecb k)\de(1-z)\theta(1-x)\,\frac{2(1-x+x^2)^2}{x}\, \bigg\{  \bigg(\frac{1}{1-x}\bigg)_+\ln \frac{p_T^2 R^2}{\mu^2} +2\bigg(\frac{\ln(1-x)}{1-x}\bigg)_+\bigg\} 
\nn \\ & \qquad
+ \de^2(\vecb k)\de(1-x)\bigg\{\theta\Big(1\geq z\geq \frac12\Big)\, \frac{2(1-z+z^2)^2}{z}\, \bigg[  \bigg(\frac{1}{1-z}\bigg)_+ \ln \frac{p_T^2 R^2z^2}{\mu^2}+2\bigg(\frac{\ln(1-z)}{1-z}\bigg)_+\bigg]
\nn \\ & \qquad
+\theta\Big(\frac12\geq z\Big)\,\frac{4(1-z+z^2)^2}{z(1-z)}\,\ln\big(z(1-z)\big)\bigg\} \bigg)
\,, \\
%%%%%%%%%%%%%%%%%%
&\cJ_{gq}^\one(x, p_T R, \vecb k, z,\mu) 
\nn \\ & \quad
= \frac{\al_s T_F}{2\pi}\,
\bigg(\frac{1}{\pi}\, \frac{1}{\mu^2}\,\frac{1}{(\vecb k^2/\mu^2)}_+\!\! \de(1-x)\,\theta\Big(\frac12 \geq z\Big)\theta\big(p_T R  \geq |\vecb k|\big)\, \big(z^2+(1-z)^2\big) 
\nn \\ & \qquad
- \de^2(\vecb k)\de(1-z)\theta(1-x) \bigg\{ \big(x^2+(1-x)^2\big) \ln \frac{p_T^2 R^2(1-x)^2}{\mu^2}  + 2x(1-x)\bigg\}
\nn \\ & \qquad
+ \de^2(\vecb k)\de(1-x) \bigg\{\theta\Big(1\geq z\geq \frac12\Big)\Big[ \big(z^2+(1-z)^2\big) \ln \frac{p_T^2 R^2z^2(1-z)^2}{\mu^2}  + 2z(1-z)\Big]\nn \\ & \qquad
+\theta\Big(\frac12\geq z\Big)\Big[2\big(z^2+(1-z)^2\big) \ln\big(z(1-z)\big)  + 2z(1-z)\Big]\bigg\} \bigg)
\,.\end{align}
%%%
The restriction $|\vecb k| \leq p_T R$ encodes the interplay between the jet boundary and $|\vecb k|$ measurement at this order. This gets ``expanded away" in \eq{match_J} when $|\vecb k| \ll p_T R$.

%===============================================================================
\subsection{TMD fragmentation function}
%===============================================================================

The matching coefficients for the universal JTMD fragmentation function in \eq{match_D} are  
%%%
\begin{align} \label{eq:C_nlo}
C_{ij}^{(0)}(\vecb k, z,\mu) &= \de_{ij}\,\de^2(\vecb k)\, \de(1-z)
\,, \\ 
%%%%%%%%%%%%%%%%%%
C_{qq}^{(1)}(\vecb k, z,\mu) &= 
\frac{\al_s C_F}{2\pi}\, \theta\Big(\frac12 \geq z\Big)
\bigg\{\frac{1}{\pi}\,\frac{1}{\mu^2} \frac{1}{(\vecb k^2/\mu^2)}_+ \! \frac{1+z^2}{1-z} 
\nn \\ & \qquad
+ \de^2(\vecb k) \bigg[\frac{2(1+z^2)}{1-z}\ln\big(z(1-z)\big) + 1-z\bigg] \bigg\}
\,, \\
%%%%%%%%%%%%%%%%%%
 C_{qg}^{(1)}(\vecb k, z,\mu) & = \frac{\al_s C_F}{2\pi}\, \theta\Big(\frac12 \geq z\Big)
\bigg\{\frac{1}{\pi}\,\frac{1}{\mu^2} \frac{1}{(\vecb k^2/\mu^2)}_+ \!\!\frac{1+(1-z)^2}{z} 
\nn \\ & \qquad 
+ \de^2(\vecb k) \Big[\frac{2(1+(1-z)^2)}{z}\ln\big(z(1-z)\big) + z\Big] \bigg\}
\,, \\
%%%%%%%%%%%%%%%%%%
 C_{gg}^{(1)}(\vecb k, z,\mu) & = 
  \frac{\al_s C_A}{2\pi}\, \theta\Big(\frac12 \geq z\Big)
\bigg\{\frac{1}{\pi}\,\frac{1}{\mu^2} \frac{1}{(\vecb k^2/\mu^2)}_+\!\!\frac{2(1-z+z^2)^2}{z(1-z)}
\nn \\ & \qquad 
+ \de^2(\vecb k)\, \frac{4(1-z+z^2)^2}{z(1-z)}\ln\big(z(1-z)\big) \bigg\}
\,, \\
%%%%%%%%%%%%%%%%%%
 C_{gq}^{(1)}(\vecb k, z,\mu) & = \frac{\al_s T_F}{2\pi}\, \theta\Big(\frac12 \geq z\Big)
\bigg\{\frac{1}{\pi}\,\frac{1}{\mu^2} \frac{1}{(\vecb k^2/\mu^2)}_+ \big(z^2+(1-z)^2\big)
\nn \\ & \qquad 
+ \de^2(\vecb k) \Big[2\big(z^2+(1-z)^2\big)\ln\big(z(1-z)\big) + 2z(1-z)\Big] \bigg\}
\,. \end{align}
%%%

%===============================================================================
\subsection{Boundary function}
%===============================================================================

The matching coefficients in \eqs{match_J}{match_G2} describe the effect of the jet boundary. They are not independent, as they can be determined from the matching coefficients $\cJ_{ij}$ and $C_{ij}$ by using \eq{match_J}. At tree level
%%%
\begin{align}
B_{ij}^\zero(x, p_T R, y,\mu) = \de_{ij}\, \de(1-x)\, \de(1-y)
\,,\end{align}
%%%
and at one-loop order, 
%%%
\begin{align}
  \cJ_{ij}^\one(x, p_T R, \vecb k, z,\mu) = 
   \Big[ \de^2(\vecb k) B_{ij}^\one(x, p_T R, z,\mu) + \de(1-x)\,C_{ij}^\one (\vecb k,z,\mu) \Big] \bigg[1 + \ORd{\frac{\vecb k^2}{p_T^2 R^2}} \bigg]
\,.\end{align}
%%%
This leads for example to
%%%
\begin{align}
B_{qq}^\one(x, p_T R, y,\mu) 
&= 
\frac{\al_s C_F}{2\pi}\,
\bigg( 
- \de(1-y)\theta(1-x) \bigg\{ (1+x^2)\Big[\Big(\frac{1}{1-x}\Big)_+\ln \Big(\frac{p_T^2 R^2}{\mu^2}\Big)
\nn \\ & \quad
+2\Big(\frac{\ln (1-x)}{1-x}\Big)_+ \Big] + 1-x \bigg\} 
+ \de(1-x) \theta\Big(1\geq y\geq \frac12\Big)
\nn \\ & \quad \times
\bigg\{ (1+y^2)\Big[\Big(\frac{1}{1-y}\Big)_+\ln \Big(\frac{p_T^2 R^2y^2}{\mu^2}\Big)+2\Big(\frac{\ln (1-y)}{1-y}\Big)_+ \Big]
 +(1-y)\bigg\} \bigg)
\,.\end{align}
%%%
The jet axis is along the most energetic of the two partons at this order. This is reflected in the expressions for the boundary functions, since they vanish for $y < 1/2$. We have also verified that the $\vecb k$-dependence cancels between $\cJ_{ij}$ and $C_{ij}$, since these boundary functions are independent of $\vecb k$.

%%%%%%%%%%%%%%%%%%%%%%%%%%%%%%%%%%%%%%%%%%%%%%%%%%%%%%%%%%%%%%%%%%%%%%%%%%%%%%%%
\section{Results for moments}
\label{sec:res}
%%%%%%%%%%%%%%%%%%%%%%%%%%%%%%%%%%%%%%%%%%%%%%%%%%%%%%%%%%%%%%%%%%%%%%%%%%%%%%%%

A full-fledged phenomenological analysis will be presented in a forthcoming publication. Here we present some first results, focussing on the transverse momentum dependence and taking moments of $z_h$. To avoid complications from distributions we integrate over the transverse momentum $|\vecb k| \leq k_c$. We will assume $p_T \gg k_c \gg \lqcd$ but not make  assumptions about the jet radius. Thus, starting from \eqs{match_D}{match_si3},
%%% 
\begin{align} \label{eq:si_mom}
  &\int_{|\mbox{\scriptsize \boldmath $k$}|< k_c} \!\!\df \vecb k\, \sum_h \int\! \df z_h\,z_h^N\,
  \frac{\df \si_h}{\df p_T\, \df \eta\, \df^2 \vecb k\, \df z_h}
  \\ 
  &\quad = \int_{|\mbox{\scriptsize \boldmath $k$}|< k_c} \!\!\df \vecb k\, \sum_h \int\! \df z_h\,z_h^N\, \sum_{i,j} \int\! \frac{\df y}{y}\, \bar \si_i(p_T, \eta, R, y, \mu) 
   \int\! \frac{\df z}{z}\, 
  C_{ij}\Big(\vecb k, \frac{z}{y},\mu\Big)\, d_j^h\Big(\frac{z_h}{z},\mu\Big)
  \nn \\ 
  &\quad = \sum_i \int\! \df y\,y^N\, \bar \si_i(p_T, \eta, R, y, \mu)\, 
  \sum_j \underbrace{\int_{|\mbox{\scriptsize \boldmath $k$}|< k_c} \!\!\df \vecb k\!\!\int\! \df z\,z^N\, C_{ij}(\vecb k, z,\mu)}_{\bar C_{ij}(k_c, N,\mu)}\, 
  \sum_h \int\! \df z_h\, z_h^N d_{j \to h}(z_h,\mu)
\,.\nn\end{align}
%%%
This implies that the transverse momentum dependence is completely governed by the matching coefficients $\bar C_{ij}(k_c, N,\mu)$, which in fixed-order perturbation theory are constant at leading order and give rise to a $\ln (k_c/\mu)$ at order $\al_s$, see \eq{C_nlo}. Note that for $N=1$ the expression in \eq{si_mom} is purely perturbative, since the dependence on the fragmentation functions drops out due to the momentum sum rule 
%%%
\begin{align}
  \sum_h \int\! \df z_h\, z_h\, d_{j \to h}(z_h,\mu) = 1
\,.\end{align}
%%%

The dependence on $k_c$ gets modified by the anomalous dimension of $\bar C$, which follows from \eqs{RGE_D}{RGE_D_k} and is multiplicative in moment space
%%%
\begin{align}\label{eq:ga_C}
  \frac{\df}{\df \ln \mu}\, \bar C_{ij}(k_c, N,\mu) &= \sum_k \big[\bar \ga_{ik}'(N,\mu) \bar C_{kj}(k_c, N,\mu) - \bar C_{ik}(k_c, N,\mu) \bar \ga_{kj}(N,\mu)\big]
\,.\end{align}
%%%
The anomalous dimensions in moment space  are given at one-loop order in \app{anom_mom}. For large values of $N$, the difference between the anomalous dimensions $\bar \ga$ and $\bar \ga'$ decreases as $2^{-N}$ and will cancel in \eq{ga_C}. Thus in that limit the transverse momentum dependence is fully captured by the fixed-order result for $\bar C$.

To diagonalize the anomalous dimension matrix in \eq{ga_C} it is convenient to perform the usual singlet/nonsinglet decomposition. Nonsinglet combinations such as $C_{qq} - C_{qq'}$ and $C_{qq} - C_{qQ}$, where $Q\neq q$ denotes a different quark flavor, do not mix. For such terms the RGE for $\bar C$ in \eq{ga_C} has as solution
%%%
\begin{align}
  U_{\rm ns}(\mu_1,\mu_0) = 
  \exp\bigg[\int_{\ln \mu_0}^{\ln \mu_1}\! \df \ln \mu\, \big(\bar \ga_{qq}'(N,\mu) - \bar \ga_{qq}(N,\mu)\big)\bigg]
\,.\end{align}
%%%
Due to the initial scale $\mu_0 \sim k_c$, which minimizes the logarithms of $k_c/\mu$ in $\bar C$, this leads to a modification of the fixed-order $k_c$ dependence in \eq{si_mom} by an additional factor
%%%
\begin{align}
k_c^{-\Delta}
\,, \qquad 
\Delta= \bar \ga_{qq}' - \bar \ga_{qq}
\,.\end{align}
%%%
At leading order $\bar C$ is independent of $k_c$, so differentiating with respect to $k_c$ to determine the $\vecb k$ dependence yields $|\vecb k|^{-2-\Delta}$.

\begin{figure}
\centering
\includegraphics[width=0.45\textwidth]{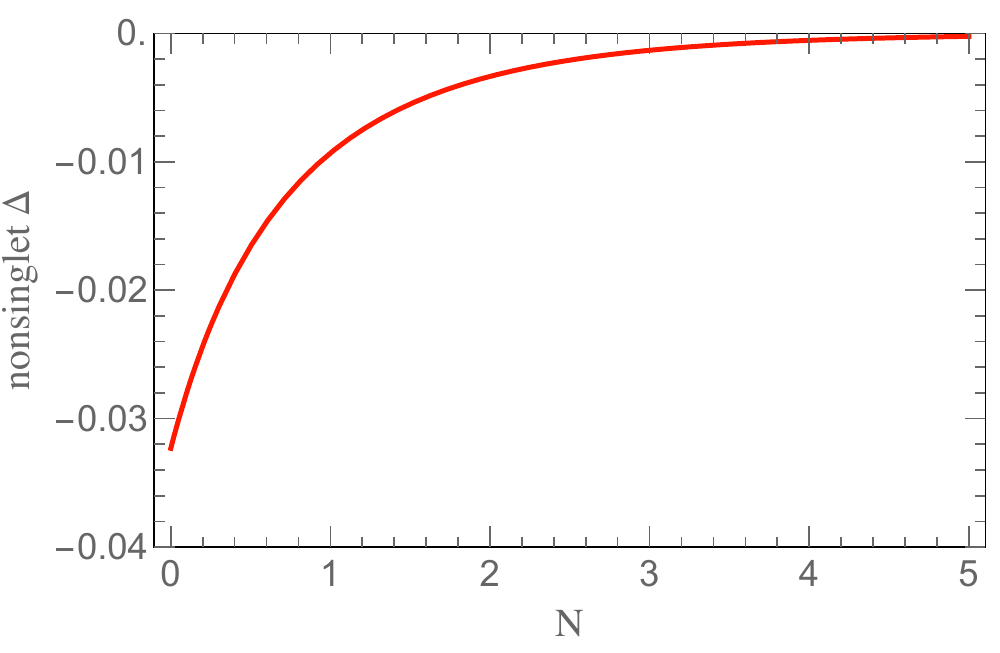}%
\qquad
\includegraphics[width=0.45\textwidth]{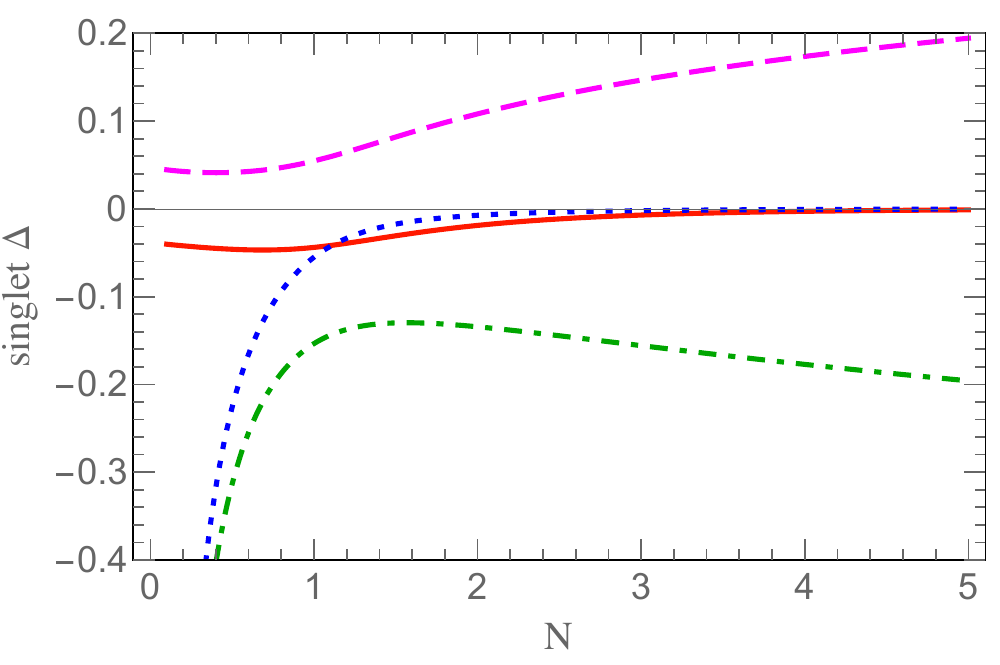}
\caption{The dependence of the cross section on the transverse momentum is given by $|\vecb k|^{-2-\Delta}$ where $\Delta$ is controlled by anomalous dimensions. The one-loop exponent $\Delta$ is shown for nonsinglet (left) and singlet (right) distributions, with $\alpha_s =0.1$. 
\label{fig:moments}}
\end{figure}

For the singlet contribution  \eq{ga_C} takes the following form
%%%
\begin{align}\label{eq:ga_sing}
  &\frac{\df}{\df \ln \mu}\, \bar C(k_c,N,\mu) = \bar \ga'(N,\mu) \bar C(k_c,N,\mu) - \bar C(k_c,N,\mu) \bar \ga(N,\mu)
  \,, \\ &
  \bar C = \begin{pmatrix} \bar C_{qq} + \bar C_{q\bar q} + (n_f\!-\!1) \bar C_{qQ} + (n_f\!-\!1) C_{q\bar Q} & \bar C_{qg} \\
   2n_f \bar  C_{gq} & \bar C_{gg} \end{pmatrix}
   , \quad
   \bar \ga' = \begin{pmatrix}
      \bar \ga_{qq}' & \bar \ga_{qg}' \\
      2 n_f \bar \ga_{gq}' & \bar  \ga_{gg}'
   \end{pmatrix}
   , \quad
   \bar \ga = \begin{pmatrix}
      \bar \ga_{qq} & \bar \ga_{qg} \\
      2 n_f \bar \ga_{gq} & \bar  \ga_{gg}
   \end{pmatrix}
.\nn\end{align}
%%%
The contributions $C_{q\bar q}$, $C_{qQ}$ and $C_{q\bar Q}$ only enter at two-loop order, but are generated by the RG evolution. There are now four different modifications $\Delta$ of the exponent of $k_c$, that can arise in a linear combination 
%%%
\begin{align} \label{eq:kc_sing}
 \sum_{i=1,\dots,4} w_i\, k_c^{-\Delta_i}
\,.\end{align}
%%%
These $\Delta_i$ are given by the differences of the eigenvalues of the anomalous dimension matrices $\bar \ga'$ and $\bar \ga$ in \eq{ga_sing}. The reason there are not two but four values is because their eigenvectors are not aligned.
Denoting the eigenvectors and eigenvalues of $\bar \ga$ by $\vec v_a$ and $\lambda_a$ and for $\bar \ga'$ by $\vec v^{\,'}_b$ and $\lambda'_b$ with $a,b=1,2$, 
%%%
\begin{align}  
   \Delta_i = \lambda'_b - \lambda_a
   \,, \qquad
   i = 2(a-1)+b
\,.\end{align}
%%%
At leading order $\bar C$ is the identity matrix. Inserting this initial condition in the RGE implies
%%%
\begin{align}
   w_i \propto \vec v^{\,'}_b \sdt \vec v_a
\,.\end{align}
%%%
For large moments $N$ the eigenvectors start to align, suggesting that two of the four weights would vanish in this limit. However, in the differential spectrum \eq{kc_sing} leads to
%%%
\begin{align} 
  \sum_{i=1,\dots,4} \Delta_i\, w_i\, |\vecb k|^{-2-\Delta_i}
,\end{align}
%%%
and these terms have a significantly larger $\Delta_i$ that compensates for their small $w_i$. The weights of course also depend on the hard scattering and fragmentation functions, and so their expressions are merely indicative. The exponents $\Delta_i$ are shown in \fig{moments} at one loop, taking $\al_s=0.1$. For the nonsinglet distributions this is probably too small to observe, but the effect should be noticeable for the singlet distributions.

%%%%%%%%%%%%%%%%%%%%%%%%%%%%%%%%%%%%%%%%%%%%%%%%%%%%%%%%%%%%%%%%%%%%%%%%%%%%%%%%
\section{Conclusions}
\label{sec:concl}
%%%%%%%%%%%%%%%%%%%%%%%%%%%%%%%%%%%%%%%%%%%%%%%%%%%%%%%%%%%%%%%%%%%%%%%%%%%%%%%%

In this paper we introduced a new definition of TMD fragmentation in jets, where the transverse momentum $\vecb k$ is measured with respect to a jet axis that is insensitive to the recoil of soft radiation. We derived factorization theorems for the regimes:
%%%
\begin{align} \label{eq:cases}
1 \gg R \gtrsim |\vecb k|/p_T
\,, \quad
1 \gg R \gg |\vecb k|/p_T
\,, \quad
1 \gtrsim R \gg |\vecb k|/p_T,
\end{align}
%%%
where $p_T$ is the jet transverse momentum and $R$ is the jet radius parameter. Angular scales that have a large hierarchy are described by different ingredients in the factorization theorem. The factorization in the latter two cases relied on the winner-take-all recombination scheme for Cambridge/Aachen or anti-$k_T$, because having a recoil-free axis was insufficient. We have calculated all the (process-independent) matching coefficients at one-loop order.

The latter two cases in \eq{cases} involve a new jet TMD fragmentation function (in the first case this cannot be separated from the jet boundary). This JTMDFF is independent of the process or the number of jets and does not involve rapidity (light-cone) divergences, because our axis choice guarantees that our observable is insensitive to soft radiation. When the transverse momentum $\vecb k$ is perturbative, the JTMD fragmentation function can be matched onto the standard fragmentation functions.

One can also consider the fragmentation of subjets instead of hadrons. One particular context where this could prove fruitful is in the area of jet substructure (see e.g.~refs.~\cite{Altheimer:2012mn,Abdesselam:2010pt,Adams:2015hiv} for an overview of developments in this field). One of the key applications of jet substructure is to identify hadronic decays of boosted heavy resonances. The boost causes the decay products to be collimated, yielding a fat jet containing subjets. Understanding the distribution of these subjets within the fat jet is critical to distinguish the desired signal from the overwhelming background of normal QCD jets. Our approach would provide analytical control over the transverse momenta (i.e.~angles) of subjets. To extend our formalism to subjets is trivial when the reclustering scale $R_{\rm sub} \ll |\vecb k|/p_T$, but requires additional calculations for other hierarchies.

The case studied in this work  treated only unpolarized  hadrons/partons.
 The angular distribution of hadrons  can  certainly be affected  by the measure of the spin and/or helicity of the  produced final state.  We postpone to a future work the study  of the sensitivity of the jet axis to the spin/helicity of final states and the relative measure of hadron spin-dependent transverse momentum. 

Another application of our framework is the study of medium effects in heavy-ion collisions. Here the modification of the momentum fraction distribution of hadrons has already been studied extensively, see e.g.~refs.~\cite{Aad:2014wha, Chatrchyan:2014ava}. Our approach would allow one to study the modification of the (relative) transverse momentum of collinear hadrons.\footnote{A perhaps more robust observable is the fragmentation of subjets in the heavy ion context.} The insensitivity of our observable to the abundant background of soft radiation present in heavy ion collisions is crucial to make this observable robust, and to be able to make meaningful comparisons to proton-proton collisions.

%%%%%%%%%%%%%%%%%%%%%%%%%%%%%%%%%%%%%%%%%%%%%%%%%%%%%%%%%%%%%%%%%%%%%%%%%%%%%%%%
\begin{acknowledgments}
We thank A.~Papaefstathiou, M.~Procura, J.~Thaler and L.~Zoppi for comments on this manuscript.
I.S.~is supported by the Spanish MECD grant FPA2014-53375-C2-2-P. W.W.~is supported by the ERC grant ERC-STG-2015-677323 and the D-ITP consortium, a program of the Netherlands Organization for Scientific Research (NWO) that is funded by the Dutch Ministry of Education, Culture and Science (OCW). D.N.~acknowledges support from US Department of Energy contract DE-AC52-06NA25396 and through the LANL/LDRD Program. We also thank the Erwin Schr\"odinger Institute program ``Challenges and Concepts for Field Theory and Applications in the Era of the LHC Run-2", where this work was initiated.
\end{acknowledgments}
%%%%%%%%%%%%%%%%%%%%%%%%%%%%%%%%%%%%%%%%%%%%%%%%%%%%%%%%%%%%%%%%%%%%%%%%%%%%%%%%

\appendix

%%%%%%%%%%%%%%%%%%%%%%%%%%%%%%%%%%%%%%%%%%%%%%%%%%%%%%%%%%%%%%%%%%%%%%%%%%%%%%%%
\section{Defining recoil-free jet functions}
\label{app:recoil}
%%%%%%%%%%%%%%%%%%%%%%%%%%%%%%%%%%%%%%%%%%%%%%%%%%%%%%%%%%%%%%%%%%%%%%%%%%%%%%%%

To make the paper as self-contained as possible, we will define the general criteria a jet function must satisfy to be recoil free, and explicitly illustrate the insensitivity to soft recoil in a one-loop example. Many different measurements can be made recoil free, and for an extensive discussion in the context of jet shapes, see ref.~\cite{Larkoski:2014uqa}. We start with a typical jet function, defined as
%%%
\begin{align}
J_n(Q,\vecb q,\tau)&=N\,\text{tr}\big\langle 0\big|\Phi_n(0)\delta(Q-\bar{n}\cdot\mathbb{P})\delta^{(2)}_\perp\big(\vecb q-\vec{\mathbb{P}}_\perp\big)\delta(\tau-\hat{O})\Phi_n(0)\big|0\big\rangle
\,,\end{align}
%%%
where $\Phi_n$ is either a quark or gluon field operator, with appropriate Wilson lines in the $\nbar$ direction for gauge invariance. $\hat{O}$ is the observable imposed on the final state of the jet function, and $\tau$ is its value. The trace is over the appropriate color and spin indicies (including the leading-power Dirac structures in the case of a quark), and $N$ normalizes the function.  We have included delta functions of the momentum operator $\mathbb{P}$ that constrain the final state of the jet function to have a total large momentum component $Q$, and a total transverse momentum $\vecb q$. The fiducial light-cone direction $n$ need not be aligned with the axis $n_\tau$ used to define the measurement $\hat{O}$. All we need is that the axis implicit in $\hat{O}$ is within a reparameterization transformation of $n$ \cite{Manohar:2002fd}. That is, if the collinear sector has assigned power counting
%%%
\begin{align}\label{eq:power_count_collinears}
p_n&\sim Q(1,\lambda^2,\lambda)
\,,\end{align}
%%%
then
\begin{align}
1-\hat{n}\cdot\hat{n}_\tau&=\ord{\lambda^2}
\,.
\end{align}
That is, the angle between $\hat{n}$ and the measurement axis is of order $\lambda$.

{\bf Definition: the jet function $J_n$ is recoil free, if the measurement of $\tau$ satisfies:}
%%%
\begin{align}\label{eq:recoil_free}
J_n(Q,\vecb q,\tau)&=J_n(Q,\vecb 0,\tau)+\mathcal{O}\bigg(\frac{|\vecb q|}{Q}\bigg)
\,,\end{align}
%%%
otherwise we call it {\bf recoil sensitive}. As an example of a recoil sensitive jet function, take the inclusive jet function found in jet mass or thrust calculations. Then $\hat{O}=n\cdot\mathbb{P}$, where $n$ is aligned with the thrust axis of the event,
%%%
\begin{align}
J_n(Q,\vecb q,\tau)&=J_n\big(Q,\vecb 0,\tau-\vecb q^{\,2}/Q\big)
\end{align}
%%%
This structure appears at all orders, and we immediately see that it fails condition \eqref{eq:recoil_free}. We can only expand out the injected transverse momentum if $\vecb q^2 \ll Q\tau$~\cite{Bauer:2008dt}.

%===============================================================================
\subsection{One-loop example}
%===============================================================================

We will now show explicitly to one-loop order that if we disturb the fiducial light-cone direction by an injection of soft recoil $\vecb q$, this has no effect on the measured transverse momentum $\vecb k$. For an all-orders discussion, see ref.~\cite{Larkoski:2014uqa}.

First we derive the form of the transverse momentum with respect to the recoil-free axis in a jet with two particles. To see that the corrections really do scale as indicated in eq. \eqref{eq:recoil_free}, we calculate the winner-take-all axis as a function of the two particle state momenta exactly, then expand in the collinear power counting. Since $k_1,k_2$ are the only two momenta in the jet, the winner-take-all axis $b$ is determined by the particle with larger energy:
%%%
\begin{align}
\text{ if } k_1^0>k_2^0: & \quad
  b(k_1,k_2)=\frac{k_1}{k_1^0}\,, \quad
  \bar{b}(k_1,k_2)=(n+\bar{n})-\frac{k_1}{k_1^0} \,,
  \nn \\
\text{ if } k_2^0>k_1^0: & \quad
  b(k_1,k_2)=\frac{k_2}{k_2^0}\,, \quad
  \bar{b}(k_1,k_2)=(n+\bar{n})-\frac{k_2}{k_2^0} \,.
\end{align}
%%%
For the conjugate $\bar b$ the sign of the spatial components is flipped, which is accomplished by the above expressions since $n+\bar{n}=(2,\vec{0})$. Then we have
%%%
\begin{align}
\text{ if } \bar{n}\cdot k_1>\bar{n}\cdot k_2: & \quad
  b(k_1,k_2)\cdot k_2=2\frac{k_1\cdot k_2}{\bar{n}\cdot k_1}\,, \quad
  \bar{b}(k_1,k_2)\cdot k_2=\bar{n}\cdot k_2
  \,, \nn \\
\text{ if } \bar{n}\cdot k_2>\bar{n}\cdot k_1: & \quad
  b(k_1,k_2)\cdot k_1=2\frac{k_1\cdot k_2}{\bar{n}\cdot k_2}\,, \quad
  \bar{b}(k_1,k_2)\cdot k_1=\bar{n}\cdot k_1\,,
\end{align}
%%%
where expanding in the collinear power counting explicitly gives corrections that scale as the small component of the momenta $k_i$ over the large momentum fraction (not the transverse scale). The relative transverse momentum of $k_1$ with respect to the winner-take-all axis is
%%%
\begin{align}
\text{ if } \bar{n}\cdot k_1>\bar{n}\cdot k_2: & \quad
 |\vecb k|=0
\,, \nn \\
\text{ if } \bar{n}\cdot k_2>\bar{n}\cdot k_1: & \quad
|\vecb k|=\frac{1}{z_h}\sqrt{b\cdot k_1\bar{b}\cdot k_1}=\frac{1}{z_h}\Big(2\frac{\nbar\cdot k_1}{\nbar\cdot k_2}k_1\cdot k_2\Big)^{1/2}+...
\end{align}
%%%

We now carry out the calculation of the one-loop JTMDFF given in \eq{Ddef}, but only to the point where we can see the independence of the recoil against the injected soft momentum. Exploiting azimuthal symmetry, we may simply consider the measurement of $\vecb k^2$. To inject soft momentum, we write the matrix element in \eq{Ddef} so that the position of the field operators acquire a transverse displacement $\vecb b_T$ as in the standard TMDFF of \eq{def_FF_opsand}. Then we take the Fourier transform at a momentum $\vecb q$ with respect to $\vecb b_T$, and integrate over the fiducial transverse momentum of the hadron. 
Taking $Q$ to be the large momentum component, the one-loop JTMDFF has the form
%%%
\begin{align} \label{eq:app_G}
D_{i\rightarrow h}^{(1)}(\vecb k,z_h;\vecb q)&=g^2 \Big(\frac{\mu^2 e^{\ga_E}}{4\pi}\Big)^\epsilon 
\int\!\frac{\df^dk_1}{(2\pi)^{d-1}}\,\theta(\bar{n}\cdot k_1)\delta(k_1^2)
\int\!\frac{\df^dk_2}{(2\pi)^{d-1}}\,\theta(\bar{n}\cdot k_2)\delta(k_2^2)
\\ &\quad\times 
(2\pi)^{d-1}\delta(Q-\bar{n}\cdot k_1-\bar{n}\cdot k_2)\delta^{(d-2)}(\vecb{k}_1+\vecb{k}_{2}-\vecb{q})\,\frac{4QC_i P_{gi}(z_h)}{(k_1+k_2)^2}\nonumber\\
&\quad\times
\delta\Big(z_h-\frac{\bar{n}\cdot k_1}{Q}\Big) \frac{1}{\pi}\bigg[\theta\Big(z_h-\frac{1}{2}\Big) \delta(\vecb k^2)+\theta\Big(\frac{1}{2}-z_h\Big) \delta\bigg(\vecb k^2-\Big(2\frac{\bar{n}\cdot k_1}{z_h^2\bar{n}\cdot k_2}\Big) k_1\cdot k_2\bigg)\bigg]
\nn \end{align}
%%%
Here we are integrating over the on-shell phase space of the two final-state partons, with momenta $k_1$ and $k_2$. The phase space is simple to interpret: The large components of the two particles sum to $Q$, while they have a non-trivial total transverse momentum $\vecb q$ with respect to the fiducial collinear direction $n$. The key point will be that the recoil-free axis is only sensitive to the relative transverse momentum of the two particles. We assume that $k_1$ is the momentum of the observed fragmented particle, which for conciseness we take to be a gluon with splitting function $P_{gi}$. The color factor $C_i$ is $C_F$ for quarks and $C_A$ for gluons.
From the delta functions in \eq{app_G} we infer:
%%%
\begin{align}
& \bar{n}\cdot k_1=Q z_h\,,\qquad\qquad \bar{n}\cdot k_2=Q(1- z_h)\,, \nn \\
& n\cdot k_1=\frac{\vecb{k}_1^2}{Q z_h}\,, \qquad\qquad n\cdot k_2=\frac{\vecb{k}_{2}^2}{Q(1- z_h)}\,, \nn \\
&\vecb{k}_{2}=\vecb{q}-\vecb{k}_{1}\,, \nn \\
&(k_1+k_2)^2=\frac{1}{z_h(1-z_h)}\big(\vecb{k}_{1}^2-2 z_h\vecb{k}_{1}\cdot\vecb{q}+z^2_h\vecb{q}^{\,2}\big) \,.
\end{align}
%%%
Performing the integrations in \eq{app_G} yields 
%%%
\begin{align}
D_{i\rightarrow h}^{(1)}(\vec{k},z_h;\vecb q) &=g^2 \Big(\frac{\mu^2 e^{\ga_E}}{4\pi}\Big)^\epsilon  C_i\int\!\frac{\df^{2-2\epsilon}\vecb{k}_{1}}{(2\pi)^{3-2\epsilon}}\,\frac{P_{gi}(z_h)}{\vecb{k}_{1}^2-2 z_h\vecb{k}_{1}\cdot\vecb{q}+z^2_h\vecb{q}^{\,2}} 
\\ &\quad \times
\frac{1}{\pi} \bigg[\theta\Big(z_h-\frac{1}{2}\Big)\delta\big(\vecb k^2\big)+\theta\Big(\frac{1}{2}-z_h\Big)\delta\bigg(\vecb k^2-\frac{\vecb{k}_{1}^2-2 z_h\vecb{k}_{1}\cdot\vecb{q}+z^2_h\vecb{q}^{\,2}}{2z_h^2(1-z_h)^2} \bigg)\bigg]
.\nn
\end{align}
%%%
We can immediately see that this function is recoil free, since the injected transverse momenta $\vecb q$ always appears in the same combination with $\vecb{k}_{1}$. Thus we can just perform a variable change and get rid of it,
%%%
\begin{align}
\vecb{k}_{1}\rightarrow\vecb{k}_{1}+z_h\vecb{q}
\,,\end{align}
%%%
making the jet function manifestly independent of $\vecb q$.

%%%%%%%%%%%%%%%%%%%%%%%%%%%%%%%%%%%%%%%%%%%%%%%%%%%%%%%%%%%%%%%%%%%%%%%%%%%%%%%%
\section{Clustering algorithms}
\label{app:cluster}
%%%%%%%%%%%%%%%%%%%%%%%%%%%%%%%%%%%%%%%%%%%%%%%%%%%%%%%%%%%%%%%%%%%%%%%%%%%%%%%%

We give a brief review of jet recombination algorithms. A more extensive discussion can be found in e.g.~ref.~\cite{Salam:2009jx}. We need a metric $d^{\alpha}(p_i,p_j;R)\equiv d_{ij}^{\alpha}(R)$ that measures the distance between two particles with momenta $p_i,p_j$ in momentum space, where $R$ is the jet radius parameter. In addition we need a single particle metric $d_{jet}^{\alpha}(p_i)\equiv d_{jet}^{\alpha}(i)$ that will decide whether a particle can be considered a jet or not. The class of metrics of interest are:
%%%
\begin{align}
&\underline{\vphantom{p}e^+e^-\text{ collision}} & & & &\underline{\text{$pp$ collision}}\nonumber\\
d^{\alpha}_{ij}(R)&=\text{min}\Big((p_i^0)^{2\alpha},(p_j^0)^{2\alpha}\Big)\frac{\theta_{ij}}{R} & & & d^{\alpha}_{ij}(R)&=\text{min}\Big(p_{Ti}^{2\alpha},p_{Tj}^{2\alpha}\Big)\frac{R_{ij}}{R}
\nn \\
d^{\alpha}_{jet}(i)&=(p_i^0)^{2\alpha} & & & d_{jet}^{\alpha}(i)&=p_{Ti}^{2\alpha}
\end{align}
%%%
In the case of $e^+e^-$ collisions, $\theta_{ij}$ is the angle between the two particles' 3-momenta, and in the case of $pp$ collisions, 
%%%
\begin{align}
R_{ij}=\sqrt{(\eta_i-\eta_j)^2+(\phi_i-\phi_j)^2}
\end{align}
%%%
is the euclidean distance between them in rapidity and azimuthal space. Note that the subscript $T$ refers to the transverse momentum with respect to the beam axis. The commonly used $k_T$ \cite{Catani:1993hr,Ellis:1993tq}, Cambridge/Aachen~\cite{Dokshitzer:1997in,Wobisch:1998wt,Wobisch:2000dk}, and anti-$k_T$~\cite{Cacciari:2008gp} algorithms correspond to  $\alpha=1,0,-1$. 

Having discussed the metrics, we now describe the algorithm. Starting with a list of particles $P=\{p_1,...,p_n\}$ with momenta $p_i$, and an empty list of jets $J=\{\}$, the recombination algorithm proceeds as follows:
\begin{enumerate}
\item If $P$ is empty, stop, output $J$. If $P$ nonempty, continue.
\item Compute $d_{ij}^{\alpha}(R)$ for all $i,j\in P$, and $d_{jet}^{\alpha}(i)$ for all $i\in P$.
\item Select the pair or the particle whose distance measure is smallest.
\item If the selection with smallest measure is a single particle, $i$, delete $p_i$ from the list $P$, move it to the list $J$.
\item If the selection with smallest measure is a pair of particles, $i,j$, delete both from $P$, {\bf merge}(i,j) them into one particle $p_{new}$, and append $P$ with $p_{new}$.
\item Go back to step 1.
\end{enumerate}
The particles inside a jet are simply all the particles that got clustered into the ``particle" that winds up in the list $J$. The {\bf merge}(i,j) procedure is usually one of the following procedures:
\begin{itemize}
\item E-scheme: $p_{new}=p_i+p_j$.
\item Winner-take-all scheme~\cite{Salam:WTAUnpublished,Bertolini:2013iqa}: writing $p_i=(p_i^0,\vec{p}_i), p_j=(p_j^0,\vec{p}_j)$, then
%%%
\begin{align}
    p^{0}_{new}&=p_i^0+p_j^0, \nn \\
    \hat{p}_{new}&=\begin{cases} 
      \frac{\vec{p}_i}{|\vec{p}_i|}\text{ if } p_i^0>p_j^0\\
      \frac{\vec{p}_j}{|\vec{p}_j|}\text{ if } p_j^0>p_i^0
      \end{cases} \nn \\
 p_{new}&=p_{new}^{0} (1,\hat{p}_{new})
\end{align}
%%%
\end{itemize}
The E-scheme results in a jet axis that aligns with the total jet momentum. Thus many properties of the thrust axis commonly used in event shape descriptions of jets also hold true for an E-scheme axis. The WTA-scheme generally has a jet axis displaced from the total jet momenta. In the case of the JTMDFF, \eq{Ddef}, we apply the clustering algorithm  assuming the final states remain in the jet. That is, we wish to only find the axis, and the jet algorithm is expanded in the limit that all particles are collinear enough, that they would always cluster before being promoted to a jet. In that case, we do not apply the single particle jet measure, and merely recombine pairwise all the particles until only one particle remains in the list $P$. That remaining particle gives the jet axis.

%%%%%%%%%%%%%%%%%%%%%%%%%%%%%%%%%%%%%%%%%%%%%%%%%%%%%%%%%%%%%%%%%%%%%%%%%%%%%%%%
\section{Results on anomalous dimensions}
\label{app:anom}
%%%%%%%%%%%%%%%%%%%%%%%%%%%%%%%%%%%%%%%%%%%%%%%%%%%%%%%%%%%%%%%%%%%%%%%%%%%%%%%%

%%%%%%%%%%%%%%%%%%%%%%%%%%%%%%%%%%%%%%%%%%%%%%%%%%%%%%%%%%%%%%%%%%%%%%%%%%%%%%%%
\subsection{All-orders anomalous dimension of JTMDFF}
%%%%%%%%%%%%%%%%%%%%%%%%%%%%%%%%%%%%%%%%%%%%%%%%%%%%%%%%%%%%%%%%%%%%%%%%%%%%%%%%

The one-loop anomalous dimension of the JTMDFF $D(\vecb k,z_h,\mu)$, defined in \eq{Ddef}, is given by
%%%
\begin{align}
  \ga_{ij}'^{\one}(z,\mu) &=\theta\Big(z \geq \frac12\Big)\, P_{ji}^\one(z)\,.
\end{align}
%%%
We will now argue this relation holds to all orders in perturbation theory, that is
%%%
\begin{align}
  \ga_{ij}'(z,\mu) &= \,\theta\Big(z \geq \frac12\Big)\sum_{\ell=1}^{\infty} P_{ji}^{(\ell)}(z)
\,,\end{align}
where $P_{ji}^{(\ell)}$ is the DGLAP splitting kernel at order $\al_s^\ell$. 
First we observe that if the parton momentum fraction $z>1/2$, the winner-take-all axis will be along its direction and $\vecb k = 0$. Thus the transverse momentum measurement does not impose a restriction on the phase space and the calculation of the JTMDFF is identical to the standard fragmentation function in this case. In particular, the IR and the UV divergences exactly match between the fragmentation function and the JTMDFF. 

For $z<1/2$ the parton can have a nontrivial transverse momentum, since the axis does not have to be aligned with it. To avoid complications from distributions, it is convenient to switch to the cumulative distribution in $\vecb k^2$. The transverse momentum of the observed parton now has an explicit upper bound due to the $\vecb k$ measurement, since for large parton transverse momenta the WTA axis will be along one of the other partons. This implies that this parton's momentum cannot scale into the UV with all the other momenta to produce a UV divergence. The only UV divergences that can occur are subdivergences corresponding to the strongly-ordered limit, which are renormalized by appropriate lower-order counter terms. 

Of course there can be new IR divergences introduced at each order, since the $\vecb k$ measurement does not prevent the transverse momenta of the partons from scaling into the IR (in a non-strongly ordered limit). Indeed, the IR divergences must exactly match those in the standard fragmentation function, including for $z<1/2$, due to \eq{match_D}. Note that we do not need to be concerned that virtual corrections will convert a $1/\epsilon_{IR}$ into a $1/\epsilon_{UV}$, since they are located at $z=1$.

%%%%%%%%%%%%%%%%%%%%%%%%%%%%%%%%%%%%%%%%%%%%%%%%%%%%%%%%%%%%%%%%%%%%%%%%%%%%%%%%
\subsection{One-loop anomalous dimensions in moment space}
\label{app:anom_mom}
%%%%%%%%%%%%%%%%%%%%%%%%%%%%%%%%%%%%%%%%%%%%%%%%%%%%%%%%%%%%%%%%%%%%%%%%%%%%%%%%

The one-loop anomalous dimensions are in moment space given by
%%%
\begin{align}
  \bar \ga_{qq}^\one(N,\mu) &= \frac{\al_s(\mu) C_F}{\pi} \Big[-2 H(N)-\frac{1}{N+1}-\frac{1}{N+2}+\frac32 \Big]
  \,, \nn \\
  \bar \ga_{qg}^\one(N,\mu) &= \frac{\al_s(\mu) C_F}{\pi} \Big[\frac{2}{N}-\frac{2}{N+1}+\frac{1}{N+2} \Big]
  \,, \nn \\
  \bar \ga_{gg}^\one(N,\mu) &= \frac{\al_s(\mu) C_A}{\pi} \Big[-2 H(N+1)+\frac{2}{N} -\frac{4}{N+1}+\frac{2}{N+2}-\frac{2}{N+3} \Big] + \frac{\al_s(\mu) \beta_0}{2\pi}
  \,, \nn \\
  \bar \ga_{gq}^\one(N,\mu) &= \frac{\al_s(\mu) T_F}{\pi} \Big[\frac{1}{N+1}-\frac{2}{N+2} + \frac{2}{N+3}\Big]
  \,, \nn \\  
  \bar \ga_{qq}'^\one(N,\mu) &= \bar \ga_{qq}^\one(N,\mu) - \frac{\al_s(\mu) C_F}{\pi}\, \big[-H_{1/2}(N)-H_{1/2}(N+2)+2 \ln 2\big]
  \,, \nn \\
  \bar \ga_{qg}'^\one(N,\mu) &=   \bar \ga_{qg}^\one(N,\mu) - \frac{\al_s(\mu) C_F}{\pi}\, 2^{-N-2}\, \frac{5N^2+17N+16}{N(N+1)(N+2)}
  \,, \nn \\
  \bar \ga_{gg}'^\one(N,\mu) &= \bar \ga_{gg}^\one(N,\mu) - \frac{\al_s(\mu) C_A}{\pi} \bigg[-2H_{1/2}(N+1) + 2\ln 2 + 2^{-N-2} \frac{5N^3+33N^2+68N+48}{N(N+1)(N+2)(N+3)} \bigg]
  \,, \nn \\
   \bar \ga_{gq}'^\one(N,\mu) &=  \bar \ga_{gq}^\one(N,\mu) - \frac{\al_s(\mu) T_F}{\pi}\, 2^{-N-2}\, \frac{N^2+5N+8}{(N+1)(N+2)(N+3)}
\,,\end{align}
%%%
where 
%%%
\begin{align}
  H(N) = \sum_{i=1}^N\, \frac{1}{i}
  \,, \qquad
  H_{1/2}(N) = \sum_{i=1}^N\, \frac{1}{i\, 2^i} = \ln 2 - 2^{-N-1} \Phi\Big(\frac12,1,N+1\Big)
\,,\end{align}
%%%
and $\Phi$ is the Lerch transcendent function. 

\phantomsection
\addcontentsline{toc}{section}{References}
\bibliographystyle{jhep}
\bibliography{TFBA}

\providecommand{\href}[2]{#2}\begingroup\raggedright\begin{thebibliography}{10}

\bibitem{Georgi:1977mg}
H.~Georgi and H.~D. Politzer, {\it {Quark Decay Functions and Heavy Hadron
  Production in QCD}},  {\em Nucl. Phys.} {\bf B136} (1978) 445--460.
%%CITATION = NUPHA,B136,445;%%

\bibitem{Ellis:1978ty}
R.~K. Ellis, H.~Georgi, M.~Machacek, H.~D. Politzer, and G.~G. Ross, {\it
  {Perturbation Theory and the Parton Model in QCD}},  {\em Nucl. Phys.} {\bf
  B152} (1979) 285--329.
%%CITATION = NUPHA,B152,285;%%

\bibitem{Collins:1981uw}
J.~C. Collins and D.~E. Soper, {\it {Parton Distribution and Decay Functions}},
   {\em Nucl. Phys.} {\bf B194} (1982) 445--492.
%%CITATION = NUPHA,B194,445;%%

\bibitem{Collins:1989gx}
J.~C. Collins, D.~E. Soper, and G.~F. Sterman, {\it {Factorization of Hard
  Processes in QCD}},  {\em Adv. Ser. Direct. High Energy Phys.} {\bf 5} (1989)
  1--91, [\href{http://arXiv.org/abs/hep-ph/0409313}{{\tt hep-ph/0409313}}].
%%CITATION = HEP-PH/0409313;%%

\bibitem{Procura:2009vm}
M.~Procura and I.~W. Stewart, {\it {Quark Fragmentation within an Identified
  Jet}},  {\em Phys. Rev.} {\bf D81} (2010) 074009,
  [\href{http://arXiv.org/abs/0911.4980}{{\tt arXiv:0911.4980}}]. [Erratum:
  Phys. Rev. D83,039902(2011)].
%%CITATION = ARXIV:0911.4980;%%

\bibitem{Liu:2010ng}
X.~Liu, {\it {SCET approach to top quark decay}},  {\em Phys. Lett.} {\bf B699}
  (2011) 87--92, [\href{http://arXiv.org/abs/1011.3872}{{\tt
  arXiv:1011.3872}}].
%%CITATION = ARXIV:1011.3872;%%

\bibitem{Jain:2011xz}
A.~Jain, M.~Procura, and W.~J. Waalewijn, {\it {Parton Fragmentation within an
  Identified Jet at NNLL}},  {\em JHEP} {\bf 05} (2011) 035,
  [\href{http://arXiv.org/abs/1101.4953}{{\tt arXiv:1101.4953}}].
%%CITATION = ARXIV:1101.4953;%%

\bibitem{Jain:2012uq}
A.~Jain, M.~Procura, B.~Shotwell, and W.~J. Waalewijn, {\it {Fragmentation with
  a Cut on Thrust: Predictions for B-factories}},  {\em Phys. Rev.} {\bf D87}
  (2013) 074013, [\href{http://arXiv.org/abs/1207.4788}{{\tt
  arXiv:1207.4788}}].
%%CITATION = ARXIV:1207.4788;%%

\bibitem{Bauer:2013bza}
C.~W. Bauer and E.~Mereghetti, {\it {Heavy Quark Fragmenting Jet Functions}},
  {\em JHEP} {\bf 04} (2014) 051, [\href{http://arXiv.org/abs/1312.5605}{{\tt
  arXiv:1312.5605}}].
%%CITATION = ARXIV:1312.5605;%%

\bibitem{Ritzmann:2014mka}
M.~Ritzmann and W.~J. Waalewijn, {\it {Fragmentation in Jets at NNLO}},  {\em
  Phys. Rev.} {\bf D90} (2014) 054029,
  [\href{http://arXiv.org/abs/1407.3272}{{\tt arXiv:1407.3272}}].
%%CITATION = ARXIV:1407.3272;%%

\bibitem{Procura:2011aq}
M.~Procura and W.~J. Waalewijn, {\it {Fragmentation in Jets: Cone and Threshold
  Effects}},  {\em Phys. Rev.} {\bf D85} (2012) 114041,
  [\href{http://arXiv.org/abs/1111.6605}{{\tt arXiv:1111.6605}}].
%%CITATION = ARXIV:1111.6605;%%

\bibitem{Chien:2015ctp}
Y.-T. Chien, Z.-B. Kang, F.~Ringer, I.~Vitev, and H.~Xing, {\it {Jet
  fragmentation functions in proton-proton collisions using soft-collinear
  effective theory}},  {\em JHEP} {\bf 05} (2016) 125,
  [\href{http://arXiv.org/abs/1512.06851}{{\tt arXiv:1512.06851}}].
%%CITATION = ARXIV:1512.06851;%%

\bibitem{Baumgart:2014upa}
M.~Baumgart, A.~K. Leibovich, T.~Mehen, and I.~Z. Rothstein, {\it {Probing
  Quarkonium Production Mechanisms with Jet Substructure}},  {\em JHEP} {\bf
  11} (2014) 003, [\href{http://arXiv.org/abs/1406.2295}{{\tt
  arXiv:1406.2295}}].
%%CITATION = ARXIV:1406.2295;%%

\bibitem{Bain:2016clc}
R.~Bain, L.~Dai, A.~Hornig, A.~K. Leibovich, Y.~Makris, and T.~Mehen, {\it
  {Analytic and Monte Carlo Studies of Jets with Heavy Mesons and Quarkonia}},
  {\em JHEP} {\bf 06} (2016) 121, [\href{http://arXiv.org/abs/1603.06981}{{\tt
  arXiv:1603.06981}}].
%%CITATION = ARXIV:1603.06981;%%

\bibitem{Arleo:2013tya}
F.~Arleo, M.~Fontannaz, J.-P. Guillet, and C.~L. Nguyen, {\it {Probing
  fragmentation functions from same-side hadron-jet momentum correlations in
  p-p collisions}},  {\em JHEP} {\bf 04} (2014) 147,
  [\href{http://arXiv.org/abs/1311.7356}{{\tt arXiv:1311.7356}}].
%%CITATION = ARXIV:1311.7356;%%

\bibitem{Kaufmann:2015hma}
T.~Kaufmann, A.~Mukherjee, and W.~Vogelsang, {\it {Hadron Fragmentation Inside
  Jets in Hadronic Collisions}},  {\em Phys. Rev.} {\bf D92} (2015) 054015,
  [\href{http://arXiv.org/abs/1506.01415}{{\tt arXiv:1506.01415}}].
%%CITATION = ARXIV:1506.01415;%%

\bibitem{Dai:2016hzf}
L.~Dai, C.~Kim, and A.~K. Leibovich, {\it {Fragmentation of a Jet with Small
  Radius}},  \href{http://arXiv.org/abs/1606.07411}{{\tt arXiv:1606.07411}}.
%%CITATION = ARXIV:1606.07411;%%

\bibitem{Kang:2016ehg}
Z.-B. Kang, F.~Ringer, and I.~Vitev, {\it {Jet substructure using
  semi-inclusive jet functions in SCET}},  {\em JHEP} {\bf 11} (2016) 155,
  [\href{http://arXiv.org/abs/1606.07063}{{\tt arXiv:1606.07063}}].
%%CITATION = ARXIV:1606.07063;%%

\bibitem{Krohn:2012fg}
D.~Krohn, M.~D. Schwartz, T.~Lin, and W.~J. Waalewijn, {\it {Jet Charge at the
  LHC}},  {\em Phys. Rev. Lett.} {\bf 110} (2013) 212001,
  [\href{http://arXiv.org/abs/1209.2421}{{\tt arXiv:1209.2421}}].
%%CITATION = ARXIV:1209.2421;%%

\bibitem{Waalewijn:2012sv}
W.~J. Waalewijn, {\it {Calculating the Charge of a Jet}},  {\em Phys. Rev.}
  {\bf D86} (2012) 094030, [\href{http://arXiv.org/abs/1209.3019}{{\tt
  arXiv:1209.3019}}].
%%CITATION = ARXIV:1209.3019;%%

\bibitem{Chang:2013rca}
H.-M. Chang, M.~Procura, J.~Thaler, and W.~J. Waalewijn, {\it {Calculating
  Track-Based Observables for the LHC}},  {\em Phys. Rev. Lett.} {\bf 111}
  (2013) 102002, [\href{http://arXiv.org/abs/1303.6637}{{\tt
  arXiv:1303.6637}}].
%%CITATION = ARXIV:1303.6637;%%

\bibitem{Chang:2013iba}
H.-M. Chang, M.~Procura, J.~Thaler, and W.~J. Waalewijn, {\it {Calculating
  Track Thrust with Track Functions}},  {\em Phys. Rev.} {\bf D88} (2013)
  034030, [\href{http://arXiv.org/abs/1306.6630}{{\tt arXiv:1306.6630}}].
%%CITATION = ARXIV:1306.6630;%%

\bibitem{Ji:2004wu}
X.-d. Ji, J.-p. Ma, and F.~Yuan, {\it {QCD factorization for semi-inclusive
  deep-inelastic scattering at low transverse momentum}},  {\em Phys. Rev.}
  {\bf D71} (2005) 034005, [\href{http://arXiv.org/abs/hep-ph/0404183}{{\tt
  hep-ph/0404183}}].
%%CITATION = HEP-PH/0404183;%%

\bibitem{Ji:2004xq}
X.-d. Ji, J.-P. Ma, and F.~Yuan, {\it {QCD factorization for spin-dependent
  cross sections in DIS and Drell-Yan processes at low transverse momentum}},
  {\em Phys. Lett.} {\bf B597} (2004) 299--308,
  [\href{http://arXiv.org/abs/hep-ph/0405085}{{\tt hep-ph/0405085}}].
%%CITATION = HEP-PH/0405085;%%

\bibitem{Becher:2010tm}
T.~Becher and M.~Neubert, {\it {{Drell-Yan} Production at Small $q_T$,
  Transverse Parton Distributions and the Collinear Anomaly}},  {\em Eur. Phys.
  J.} {\bf C71} (2011) 1665, [\href{http://arXiv.org/abs/1007.4005}{{\tt
  arXiv:1007.4005}}].
%%CITATION = ARXIV:1007.4005;%%

\bibitem{Collins:2011zzd}
J.~Collins, {\em {Foundations of perturbative QCD}}.
\newblock Cambridge University Press, 2013.
%%CITATION = INSPIRE-922696;%%

\bibitem{Chiu:2012ir}
J.-Y. Chiu, A.~Jain, D.~Neill, and I.~Z. Rothstein, {\it {A Formalism for the
  Systematic Treatment of Rapidity Logarithms in Quantum Field Theory}},  {\em
  JHEP} {\bf 1205} (2012) 084, [\href{http://arXiv.org/abs/1202.0814}{{\tt
  arXiv:1202.0814}}].
%%CITATION = ARXIV:1202.0814;%%

\bibitem{GarciaEchevarria:2011rb}
M.~G. Echevarria, A.~Idilbi, and I.~Scimemi, {\it {Factorization Theorem For
  Drell-Yan At Low $q_T$ And Transverse Momentum Distributions
  On-The-Light-Cone}},  {\em JHEP} {\bf 07} (2012) 002,
  [\href{http://arXiv.org/abs/1111.4996}{{\tt arXiv:1111.4996}}].
%%CITATION = ARXIV:1111.4996;%%

\bibitem{Bertolini:2013iqa}
D.~Bertolini, T.~Chan, and J.~Thaler, {\it {Jet Observables Without Jet
  Algorithms}},  {\em JHEP} {\bf 04} (2014) 013,
  [\href{http://arXiv.org/abs/1310.7584}{{\tt arXiv:1310.7584}}].
%%CITATION = ARXIV:1310.7584;%%

\bibitem{Larkoski:2014uqa}
A.~J. Larkoski, D.~Neill, and J.~Thaler, {\it {Jet Shapes with the Broadening
  Axis}},  {\em JHEP} {\bf 04} (2014) 017,
  [\href{http://arXiv.org/abs/1401.2158}{{\tt arXiv:1401.2158}}].
%%CITATION = ARXIV:1401.2158;%%

\bibitem{Salam:WTAUnpublished}
G.~Salam, {\it {$E_t^\infty$ Scheme}},  {\em Unpublished}.

\bibitem{Ellis:2010rwa}
S.~D. Ellis, C.~K. Vermilion, J.~R. Walsh, A.~Hornig, and C.~Lee, {\it {Jet
  Shapes and Jet Algorithms in SCET}},  {\em JHEP} {\bf 11} (2010) 101,
  [\href{http://arXiv.org/abs/1001.0014}{{\tt arXiv:1001.0014}}].
%%CITATION = ARXIV:1001.0014;%%

\bibitem{Dasgupta:2001sh}
M.~Dasgupta and G.~P. Salam, {\it {Resummation of nonglobal QCD observables}},
  {\em Phys. Lett.} {\bf B512} (2001) 323--330,
  [\href{http://arXiv.org/abs/hep-ph/0104277}{{\tt hep-ph/0104277}}].
%%CITATION = HEP-PH/0104277;%%

\bibitem{Stewart:2010tn}
I.~W. Stewart, F.~J. Tackmann, and W.~J. Waalewijn, {\it {N-Jettiness: An
  Inclusive Event Shape to Veto Jets}},  {\em Phys. Rev. Lett.} {\bf 105}
  (2010) 092002, [\href{http://arXiv.org/abs/1004.2489}{{\tt
  arXiv:1004.2489}}].
%%CITATION = ARXIV:1004.2489;%%

\bibitem{Forshaw:2012bi}
J.~R. Forshaw, M.~H. Seymour, and A.~Siodmok, {\it {On the Breaking of
  Collinear Factorization in QCD}},  {\em JHEP} {\bf 11} (2012) 066,
  [\href{http://arXiv.org/abs/1206.6363}{{\tt arXiv:1206.6363}}].
%%CITATION = ARXIV:1206.6363;%%

\bibitem{Rogers:2010dm}
T.~C. Rogers and P.~J. Mulders, {\it {No Generalized TMD-Factorization in
  Hadro-Production of High Transverse Momentum Hadrons}},  {\em Phys. Rev.}
  {\bf D81} (2010) 094006, [\href{http://arXiv.org/abs/1001.2977}{{\tt
  arXiv:1001.2977}}].
%%CITATION = ARXIV:1001.2977;%%

\bibitem{Catani:2011st}
S.~Catani, D.~de~Florian, and G.~Rodrigo, {\it {Space-like (versus time-like)
  collinear limits in QCD: Is factorization violated?}},  {\em JHEP} {\bf 07}
  (2012) 026, [\href{http://arXiv.org/abs/1112.4405}{{\tt arXiv:1112.4405}}].
%%CITATION = ARXIV:1112.4405;%%

\bibitem{Gaunt:2014ska}
J.~R. Gaunt, {\it {Glauber Gluons and Multiple Parton Interactions}},  {\em
  JHEP} {\bf 07} (2014) 110, [\href{http://arXiv.org/abs/1405.2080}{{\tt
  arXiv:1405.2080}}].
%%CITATION = ARXIV:1405.2080;%%

\bibitem{Zeng:2015iba}
M.~Zeng, {\it {Drell-Yan process with jet vetoes: breaking of generalized
  factorization}},  {\em JHEP} {\bf 10} (2015) 189,
  [\href{http://arXiv.org/abs/1507.01652}{{\tt arXiv:1507.01652}}].
%%CITATION = ARXIV:1507.01652;%%

\bibitem{Rothstein:2016bsq}
I.~Z. Rothstein and I.~W. Stewart, {\it {An Effective Field Theory for Forward
  Scattering and Factorization Violation}},  {\em JHEP} {\bf 08} (2016) 025,
  [\href{http://arXiv.org/abs/1601.04695}{{\tt arXiv:1601.04695}}].
%%CITATION = ARXIV:1601.04695;%%

\bibitem{Bain:2016rrv}
R.~Bain, Y.~Makris, and T.~Mehen, {\it {Transverse Momentum Dependent
  Fragmenting Jet Functions with Applications to Quarkonium Production}},
  \href{http://arXiv.org/abs/1610.06508}{{\tt arXiv:1610.06508}}.
%%CITATION = ARXIV:1610.06508;%%

\bibitem{Kelley:2011aa}
R.~Kelley, M.~D. Schwartz, R.~M. Schabinger, and H.~X. Zhu, {\it {Jet Mass with
  a Jet Veto at Two Loops and the Universality of Non-Global Structure}},  {\em
  Phys. Rev.} {\bf D86} (2012) 054017,
  [\href{http://arXiv.org/abs/1112.3343}{{\tt arXiv:1112.3343}}].
%%CITATION = ARXIV:1112.3343;%%

\bibitem{Bauer:2011uc}
C.~W. Bauer, F.~J. Tackmann, J.~R. Walsh, and S.~Zuberi, {\it {Factorization
  and Resummation for Dijet Invariant Mass Spectra}},  {\em Phys. Rev.} {\bf
  D85} (2012) 074006, [\href{http://arXiv.org/abs/1106.6047}{{\tt
  arXiv:1106.6047}}].
%%CITATION = ARXIV:1106.6047;%%

\bibitem{Larkoski:2015zka}
A.~J. Larkoski, I.~Moult, and D.~Neill, {\it {Non-Global Logarithms,
  Factorization, and the Soft Substructure of Jets}},  {\em JHEP} {\bf 09}
  (2015) 143, [\href{http://arXiv.org/abs/1501.04596}{{\tt arXiv:1501.04596}}].
%%CITATION = ARXIV:1501.04596;%%

\bibitem{Pietrulewicz:2016nwo}
P.~Pietrulewicz, F.~J. Tackmann, and W.~J. Waalewijn, {\it {Factorization and
  Resummation for Generic Hierarchies between Jets}},  {\em JHEP} {\bf 08}
  (2016) 002, [\href{http://arXiv.org/abs/1601.05088}{{\tt arXiv:1601.05088}}].
%%CITATION = ARXIV:1601.05088;%%

\bibitem{Bauer:2000ew}
C.~W. Bauer, S.~Fleming, and M.~E. Luke, {\it {Summing Sudakov logarithms in B
  $\to$ X$_s \gamma$ in effective field theory}},  {\em Phys. Rev.} {\bf D63}
  (2000) 014006, [\href{http://arXiv.org/abs/hep-ph/0005275}{{\tt
  hep-ph/0005275}}].
%%CITATION = HEP-PH/0005275;%%

\bibitem{Bauer:2000yr}
C.~W. Bauer, S.~Fleming, D.~Pirjol, and I.~W. Stewart, {\it {An Effective field
  theory for collinear and soft gluons: Heavy to light decays}},  {\em Phys.
  Rev.} {\bf D63} (2001) 114020,
  [\href{http://arXiv.org/abs/hep-ph/0011336}{{\tt hep-ph/0011336}}].
%%CITATION = HEP-PH/0011336;%%

\bibitem{Bauer:2001ct}
C.~W. Bauer and I.~W. Stewart, {\it {Invariant operators in collinear effective
  theory}},  {\em Phys. Lett.} {\bf B516} (2001) 134--142,
  [\href{http://arXiv.org/abs/hep-ph/0107001}{{\tt hep-ph/0107001}}].
%%CITATION = HEP-PH/0107001;%%

\bibitem{Bauer:2001yt}
C.~W. Bauer, D.~Pirjol, and I.~W. Stewart, {\it {Soft collinear factorization
  in effective field theory}},  {\em Phys. Rev.} {\bf D65} (2002) 054022,
  [\href{http://arXiv.org/abs/hep-ph/0109045}{{\tt hep-ph/0109045}}].
%%CITATION = HEP-PH/0109045;%%

\bibitem{Echevarria:2016scs}
M.~G. Echevarria, I.~Scimemi, and A.~Vladimirov, {\it {Unpolarized Transverse
  Momentum Dependent Parton Distribution and Fragmentation Functions at
  next-to-next-to-leading order}},  \href{http://arXiv.org/abs/1604.07869}{{\tt
  arXiv:1604.07869}}.
%%CITATION = ARXIV:1604.07869;%%

\bibitem{Idilbi:2010im}
A.~Idilbi and I.~Scimemi, {\it {Singular and Regular Gauges in Soft Collinear
  Effective Theory: The Introduction of the New Wilson Line T}},  {\em Phys.
  Lett.} {\bf B695} (2011) 463--468,
  [\href{http://arXiv.org/abs/1009.2776}{{\tt arXiv:1009.2776}}].
%%CITATION = ARXIV:1009.2776;%%

\bibitem{GarciaEchevarria:2011md}
M.~Garcia-Echevarria, A.~Idilbi, and I.~Scimemi, {\it {SCET, Light-Cone Gauge
  and the T-Wilson Lines}},  {\em Phys. Rev.} {\bf D84} (2011) 011502,
  [\href{http://arXiv.org/abs/1104.0686}{{\tt arXiv:1104.0686}}].
%%CITATION = ARXIV:1104.0686;%%

\bibitem{Manohar:2002fd}
A.~V. Manohar, T.~Mehen, D.~Pirjol, and I.~W. Stewart, {\it {Reparameterization
  invariance for collinear operators}},  {\em Phys. Lett.} {\bf B539} (2002)
  59--66, [\href{http://arXiv.org/abs/hep-ph/0204229}{{\tt hep-ph/0204229}}].
%%CITATION = HEP-PH/0204229;%%

\bibitem{Dokshitzer:1997in}
Y.~L. Dokshitzer, G.~D. Leder, S.~Moretti, and B.~R. Webber, {\it {Better jet
  clustering algorithms}},  {\em JHEP} {\bf 08} (1997) 001,
  [\href{http://arXiv.org/abs/hep-ph/9707323}{{\tt hep-ph/9707323}}].
%%CITATION = HEP-PH/9707323;%%

\bibitem{Wobisch:1998wt}
M.~Wobisch and T.~Wengler, {\it {Hadronization corrections to jet
  cross-sections in deep inelastic scattering}},
  \href{http://arXiv.org/abs/hep-ph/9907280}{{\tt hep-ph/9907280}}.
%%CITATION = HEP-PH/9907280;%%

\bibitem{Wobisch:2000dk}
M.~Wobisch, {\it {Measurement and QCD analysis of jet cross-sections in deep
  inelastic positron proton collisions at $\sqrt{s} = 300$~GeV}}, .
%%CITATION = DESY-THESIS-2000-049 ETC.;%%

\bibitem{Cacciari:2008gp}
M.~Cacciari, G.~P. Salam, and G.~Soyez, {\it {The Anti-k$_t$ jet clustering
  algorithm}},  {\em JHEP} {\bf 0804} (2008) 063,
  [\href{http://arXiv.org/abs/0802.1189}{{\tt arXiv:0802.1189}}].
%%CITATION = ARXIV:0802.1189;%%

\bibitem{Luke:1992cs}
M.~E. Luke and A.~V. Manohar, {\it {Reparametrization invariance constraints on
  heavy particle effective field theories}},  {\em Phys. Lett.} {\bf B286}
  (1992) 348--354, [\href{http://arXiv.org/abs/hep-ph/9205228}{{\tt
  hep-ph/9205228}}].
%%CITATION = HEP-PH/9205228;%%

\bibitem{Bauer:2008dt}
C.~W. Bauer, S.~P. Fleming, C.~Lee, and G.~F. Sterman, {\it {Factorization of
  e+e- Event Shape Distributions with Hadronic Final States in Soft Collinear
  Effective Theory}},  {\em Phys. Rev.} {\bf D78} (2008) 034027,
  [\href{http://arXiv.org/abs/0801.4569}{{\tt arXiv:0801.4569}}].
%%CITATION = ARXIV:0801.4569;%%

\bibitem{Kang:2016mcy}
Z.-B. Kang, F.~Ringer, and I.~Vitev, {\it {The semi-inclusive jet function in
  SCET and small radius resummation for inclusive jet production}},
  \href{http://arXiv.org/abs/1606.06732}{{\tt arXiv:1606.06732}}.
%%CITATION = ARXIV:1606.06732;%%

\bibitem{Gribov:1972ri}
V.~N. Gribov and L.~N. Lipatov, {\it {Deep inelastic e p scattering in
  perturbation theory}},  {\em Sov. J. Nucl. Phys.} {\bf 15} (1972) 438--450.
  [Yad. Fiz.15,781(1972)].
%%CITATION = SJNCA,15,438;%%

\bibitem{Altarelli:1977zs}
G.~Altarelli and G.~Parisi, {\it {Asymptotic Freedom in Parton Language}},
  {\em Nucl. Phys.} {\bf B126} (1977) 298--318.
%%CITATION = NUPHA,B126,298;%%

\bibitem{Dokshitzer:1977sg}
Y.~L. Dokshitzer, {\it {Calculation of the Structure Functions for Deep
  Inelastic Scattering and e+ e- Annihilation by Perturbation Theory in Quantum
  Chromodynamics.}},  {\em Sov. Phys. JETP} {\bf 46} (1977) 641--653. [Zh.
  Eksp. Teor. Fiz.73,1216(1977)].
%%CITATION = SPHJA,46,641;%%

\bibitem{Altheimer:2012mn}
A.~Altheimer {\em et~al.}, {\it {Jet Substructure at the Tevatron and LHC: New
  results, new tools, new benchmarks}},  {\em J. Phys.} {\bf G39} (2012)
  063001, [\href{http://arXiv.org/abs/1201.0008}{{\tt arXiv:1201.0008}}].
%%CITATION = ARXIV:1201.0008;%%

\bibitem{Abdesselam:2010pt}
A.~Abdesselam {\em et~al.}, {\it {Boosted objects: A Probe of beyond the
  Standard Model physics}},  {\em Eur. Phys. J.} {\bf C71} (2011) 1661,
  [\href{http://arXiv.org/abs/1012.5412}{{\tt arXiv:1012.5412}}].
%%CITATION = ARXIV:1012.5412;%%

\bibitem{Adams:2015hiv}
D.~Adams {\em et~al.}, {\it {Towards an Understanding of the Correlations in
  Jet Substructure}},  {\em Eur. Phys. J.} {\bf C75} (2015) 409,
  [\href{http://arXiv.org/abs/1504.00679}{{\tt arXiv:1504.00679}}].
%%CITATION = ARXIV:1504.00679;%%

\bibitem{Aad:2014wha}
{\bf ATLAS} Collaboration, G.~Aad {\em et~al.}, {\it {Measurement of inclusive
  jet charged-particle fragmentation functions in Pb+Pb collisions at
  $\sqrt{s_{NN}}$=2.76 TeV with the ATLAS detector}},  {\em Phys. Lett.} {\bf
  B739} (2014) 320--342, [\href{http://arXiv.org/abs/1406.2979}{{\tt
  arXiv:1406.2979}}].
%%CITATION = ARXIV:1406.2979;%%

\bibitem{Chatrchyan:2014ava}
{\bf CMS} Collaboration, S.~Chatrchyan {\em et~al.}, {\it {Measurement of jet
  fragmentation in PbPb and pp collisions at $\sqrt{s_{NN}}=2.76$ TeV}},  {\em
  Phys. Rev.} {\bf C90} (2014) 024908,
  [\href{http://arXiv.org/abs/1406.0932}{{\tt arXiv:1406.0932}}].
%%CITATION = ARXIV:1406.0932;%%

\bibitem{Salam:2009jx}
G.~P. Salam, {\it {Towards Jetography}},  {\em Eur. Phys. J.} {\bf C67} (2010)
  637--686, [\href{http://arXiv.org/abs/0906.1833}{{\tt arXiv:0906.1833}}].
%%CITATION = ARXIV:0906.1833;%%

\bibitem{Catani:1993hr}
S.~Catani, Y.~L. Dokshitzer, M.~Seymour, and B.~Webber, {\it {Longitudinally
  invariant $K_t$ clustering algorithms for hadron hadron collisions}},  {\em
  Nucl. Phys.} {\bf B406} (1993) 187--224.
%%CITATION = NUPHA,B406,187;%%

\bibitem{Ellis:1993tq}
S.~D. Ellis and D.~E. Soper, {\it {Successive combination jet algorithm for
  hadron collisions}},  {\em Phys. Rev.} {\bf D48} (1993) 3160--3166,
  [\href{http://arXiv.org/abs/hep-ph/9305266}{{\tt hep-ph/9305266}}].
%%CITATION = HEP-PH/9305266;%%

\end{thebibliography}\endgroup

\end{document}